\documentclass[preprint,12pt]{elsarticle}

\usepackage{amssymb}
\usepackage{amsthm}
\usepackage{xcolor}
\usepackage{graphicx}
\usepackage{lineno,hyperref}
\modulolinenumbers[5]

\usepackage{inputenc}
\usepackage{amsmath}
\usepackage{graphics}
\usepackage{epsfig}
\usepackage{graphicx}
\usepackage{latexsym}
\usepackage{graphics}
\usepackage{epsfig}
\usepackage{amsmath,amssymb,amsfonts,theorem,color,bm}
\usepackage{multirow}
\usepackage{stfloats} 
\usepackage{float}
\usepackage{subfig}
\usepackage{soul}
\usepackage{caption}
\usepackage{hyperref}
\hypersetup{colorlinks,allcolors=black}

\newcommand\bv{\boldsymbol v}
\newcommand\bV{\boldsymbol V}

\newcommand\bW{\boldsymbol W}

\newcommand\bT{\boldsymbol T}

\newcommand\bSigma{\boldsymbol{\Sigma}}

\numberwithin{equation}{section}

\newcommand{\beqn}{\begin{equation}}
\newcommand{\eeqn}{\end{equation}}
\newcommand{\beqnarr}{\begin{eqnarray}}
\newcommand{\eeqnarr}{\end{eqnarray}}
\newcommand{\baling}{\begin{alignat}{1}}
\newcommand{\ealing}{\end{alignat}}

\usepackage{xcolor,colortbl}
\definecolor{Gray}{gray}{0.75}
\newcolumntype{a}{>{\columncolor{Gray}}c}

\journal{XXX}

\begin{document}

\begin{frontmatter}



\title{Data repairing and resolution enhancement using data-driven modal decomposition and deep learning}


\author[UPM]{Ashton Hetherington}

\author[TUM]{Daniel Serfaty Bayón}

\author[UPM]{Adrián Corrochano}

\author[MONASH]{Julio Soria}

\author[UPM]{Soledad Le Clainche\footnote{Correspondence to: soledad.leclainche@upm.es}}

\affiliation[UPM]{organization={ETSI Aeronáutica y del Espacio, Universidad Politécnica de Madrid},
            addressline={Plaza Cardenal Cisneros, 3}, 
            city={Madrid},
            postcode={28040}, 
            country={Spain}}

\affiliation[TUM]{organization={Mechanical and Aerospace Engineering Faculty, Technische Universität München},
            addressline={Boltzmannstraße 15}, 
            city={Garching},
            postcode={85748}, 
            country={Germany}}

\affiliation[MONASH]{organization={Laboratory for Turbulence Research in Aerospace and Combustion (LTRAC), Department of 
Mechanical and Aerospace Engineering, Monash University}, 
            city={Melbourne VIC 3800},
            country={Australia}}
      
\begin{abstract}
This paper introduces a new series of methods which combine modal decomposition algorithms, such as singular value decomposition and high-order singular value decomposition, and deep learning architectures to repair, enhance, and increase the quality and precision of numerical and experimental data. A combination of two- and three-dimensional, numerical and experimental dasasets are used to demonstrate the reconstruction capacity of the presented methods, showing that these methods can be used to reconstruct any type of dataset, showing outstanding results when applied to highly complex data, which is noisy. The combination of benefits of these techniques results in a series of data-driven methods which are capable of repairing and/or enhancing the resolution of a dataset by identifying the underlying physics that define the data, which is incomplete or under-resolved, filtering any existing noise. These methods and the \emph{Python} codes are included in the first release of ModelFLOWs-app\footnote{The website of the software is available at \href{https://modelflows.github.io/modelflowsapp/}{https://modelflows.github.io/modelflowsapp/}}.
\end{abstract}




\end{frontmatter}



\section{Introduction\label{sec:introduction}}

In fluid mechanics, it is common to conduct experiments that typically suffer from low-resolution or employ sensors that provide limited information due to their low spatial resolution. Moreover, experimental data always contains some level of noise, a significant factor impacting both data quality and resolution. Clarity and completeness are crucial when describing the behavior of fluid motions based on experimental data. Concerning sensor usage, challenges may arise during data collection. Sensor failures during experiments can result in incomplete datasets, particularly with missing data corresponding to the positions of faulty sensors. This issue can lead to research delays, necessitating experiment repetition or postponement until damaged sensors are repaired, incurring additional costs. Another prevalent problem involves insufficient resources for experiments, such as an inadequate number of sensors to capture all the dynamics of the studied system \cite{erichson2020shallow}. This deficiency results in data with reduced resolution and, consequently, lower quality.

On the other hand, computational fluid dynamics (CFD) simulations offer an excellent alternative to experiments, providing comprehensive and highly detailed data. However, a notable drawback of numerical simulations is their high computational cost, particularly when tackling realistic scenarios or solving industrial problems involving turbulence. The ability to conduct low-resolution numerical simulations and subsequently enhance data quality during postprocessing without sacrificing precision would yield significant computational cost benefits. This approach enables effective solutions for various complex simulations. Challenges also arise in numerical simulations using computational fluid dynamics software. While these tools can generate highly detailed datasets describing the interaction between a fluid and a body, they come with substantial computational costs. An alternative, avoiding the need to acquire additional resources, involves generating low-resolution numerical simulation data \cite{roelofs2012simulating} and then improving the resolution using a postprocessing tool.
There is a large variety of fields where both experiments and simulations are performed to solve complex fluid dynamics problems. In the aerospace industry, experiments and simulations are performed to analyze aircraft behaviour \citep{zyskowski2003aircraft} in order to create new highly efficient designs that are less pollutant. Another example is the analysis of flutter instability in flight test experiments \cite{mendez2021new}, which is important to study the structural integrity of an aircraft. Another relevant field is combustion, where data is collected to study reactive flows \cite{paperAdricombustion}, just to name one of many applications. In the medical field, fluid dynamics data is used for modelling within cardiovascular medicine \cite{morris2016computational}, with the main goal being to detect cardiovascular diseases. A field that continues to gain relevancy is urban flows, where highly precise data is essential in order to understand the dispersion of pollutants in urban environments \citep{torres2021experimental, Torresetal}. The data used to study air quality is collected by sensor networks which are distributed among some of the most important cities in the world. These are just some of the many fields where high detailed fluid dynamics data plays an important role.

In industrial settings, a substantial portion of time and financial resources is allocated towards the development and simulation phases of diverse models, particularly evident in the aerospace industry \citep{LeClaincheEnergies19}. The utilization of reconstruction algorithms characterized by both cost-effectiveness and high precision presents an avenue for economizing resources in complex simulations and optimizing time management. Therefore, there is rising necessity to find new techniques capable of restoring damaged, incomplete or under-resolved datasets, repairing the data \citep{ruscher2017repairing} or enhancing the data resolution \citep{vinuesa2022enhancing} without the need for repeating experimental setups or performing high computational cost simulations \citep{brunton2020machine}.

This article presents a comparison between the reconstruction results obtained using a series of data reparation and resolution enhancements methods based on physics principles, which use modal decomposition techniques, such as singular value decomposition, and deep learning neural networks. The main objective is to demonstrate the main differences that exist between these methods when applied to data varying in complexity, such as two- or three-dimensional data, which can be laminar or turbulent, obtained using numerical software or performing experiments. Large emphasis is placed on testing these methods on experimental data, since they have never been applied to this type of data before, challenging their reconstruction capabilities and limitations.

Before explaining these methodologies, it is important to clarify some fundamental terms which are repeated throughout this article. The term \textit{data reconstruction} is used as a general expression to describe all processes in which initially generated data, which may apparently seem vain, is transformed in a way that its information can be used effectively. This can be achieved by using \textit{data repairing} methods, which are used to fill in the missing data from a dataset, or via \textit{data resolution enhancement} methods, where, starting from a reduced number of data points, high resolution data is generated by creating new data based on that which is available. 

There is preexisting work in this line of research, where modal decomposition or deep learning have been used to reconstruct data. Modal decomposition techniques, in particular Singular Value Decomposition (SVD), have been used to enhance data resolution, such as in \cite{downs2021resolution}, where it is used to enhance the resolution of deconvolved ground penetrating radar images, in \citep{lenti2013two}, with its use to enhance the spatial resolution of radiometer data, in \citep{shamna2014satellite}, SVD is used for satellite image resolution and brightness enhancement. SVD has also been used to repair missing data, as in Refs. \citep{intawichai2022missing, chen2018spatial}.

Deep learning has also proven to have great potential with data reconstruction, \citep{yousif2023deep}, where a generative adversarial network (GAN) is designed to reconstruct three-dimensional velocity fields, or in \citep{guemes2022super}, where a neural network named RAndomly-SEEDed super-resolution GAN (RaSeedGAN) is proven capable of transforming a low-resolution image to a high-resolution image. Even though GANs have gained a large popularity in recent times given their outstanding image generation capabilities, other architectures, such as convolutional neural networks and autoencoders can also be used to reconstruct dataset, \citep{bolton2019applications, fukami2019super, AumAE, AumVAE}.

The reconstruction methods described in this work combine the benefits of modal decomposition and deep learning, creating hybrid reduced order models (ROMs). Modal decomposition techniques have proven to be useful in fluid dynamics \citep{VegaLeClaincheBook20}, given their good capabilities of identifying and extracting the most important patterns, representing the flow physics, contained in a dataset \citep{Corrochano22, Lazpita}. On the other hand, deep learning neural networks have proven to have a high capability of learning complex patterns from reduced data. For these reasons, a series of methodologies, which can be purely based on modal decomposition or hybrid, combining modal decomposition and deep learning, have been developed.

The main difference between the methods presented in this article versus the preexisting methods is the fusion of the previously explained techniques to create powerful, robust tools that combine the benefits of each of these to reconstruct data with high precision and a low computational cost. Contrary to classic methods, these new methods perform data reconstruction based on the information of underlying physics contained in the data which explains the fluid behaviour. This information can then be used to complete missing data or perform a high resolution reconstruction. In other words, the presented methodologies understand the information contained in the dataset that is being reconstructed. Additionally, with the use of neural networks, temporal predictions of high resolution data can be generated using preexisting low resolution data in order to obtain new data. The development of a these new, simple and powerful dataset reconstruction methods presents a variety benefits when implemented in an extended range of fields, such as those described at the beginning of this section.

This work is organized as follows. Section \ref{sec:methodology} offers a detailed explanation of the methodology of these novel algorithms, which are used to repair datasets or enhance their resolution. These new methods are tested by applying them to a series of test cases, which consist of a variety of datasets, containing two- and three-dimensional, laminar and turbulent, numerical and experimental data, and are presented in Sec. \ref{sec:database}. The results of the application of the reconstruction algorithms to these test case datasets are shown in Sec. \ref{sec:results}, where a variety of case results are shown for each test case dataset and application. Finally, the conclusions extracted from this work are gathered in Sec. \ref{sec:conclusions}, where the differences between these different algorithms are discussed.

\section{Data reparation and resolution enhancement based on modal decomposition methods and deep learning\label{sec:methodology}}

Modal decomposition and machine learning methods have been combined to create a series of data-driven algorithms and hybrid ROMs capable of reconstructing datasets, whether it is to repair missing or corrupt data, or to enhance the resolution of a dataset. This section presents an overview of each method, with more in-depth information about each one of these being available in \cite{hetherington2023modelflows}.

\subsection{Data organization \label{sec:dataOrganization}}

The following fully data-driven algorithms require datasets to be in matrix or tensor form in order to be correctly reconstructed.  

When a dataset is organized in matrix form, it consists of a set of $K$ snapshots $\bv_k=\bv(t_k)$, with $t_k$ being the time measurement that corresponds to each snapshot $k$. A dataset in matrix form, from now on a \textbf{snapshot matrix}, is organized as follows 

\beqn
\bV_1^K = [\bv_{1},\bv_{2},\ldots,\bv_{k},\bv_{k+1},\ldots,\bv_{K-1},\bv_{K}].\label{ab0}
\eeqn

In some cases, it is more convenient to re-organize the data into tensor form, referred to as a \textit{snapshot tensor}. Snapshot tensors use multidimensional indexing, with \textit{fibers} being one-dimensional slices representing snapshot matrices. This structure is ideal for storing and analyzing complex data. 
The algorithms described in this paper generally used two- and three-dimensional data which is structured as forth and fifth order tensors, respectively. 

For instance, three-dimensional data is organized into a fifth-order snapshot tensor $\bV$, with dimensions $J_1\times J_2\times J_3\times J_4\times K$, whose components $V_{j_1j_2j_3j_4k}$ can be defined as 
 \beqn
 \begin{split}
V_{1j_2j_3k}&=v_1(x_{j_2},y_{j_3},z_{j_4},t_k),\\ V_{2j_2j_3k}&=v_2(x_{j_2},y_{j_3},z_{j_4},t_k),\\
V_{3j_2j_3k}&=v_2(x_{j_2},y_{j_3},z_{j_4},t_k),\label{c43d}
\end{split}
\eeqn
%
where $J_1$ represents the number of components of the database, which in this case are the three velocity components: the streamwise, normal, and spanwise velocities. Indexes $j_2$, $j_3$ and $j_4$ correspond to the discrete values of the three spatial coordinates, $x$, $y$ and $z$, and $k$ represents the time instant.

A correlation exists between snapshot matrices and tensors since each set of data collected for every individual time instant of a snapshot tensor can be compressed into a single dimension, forming a snapshot matrix. Therefore, the snapshot tensor indices $j_1$, $j_2$, $j_3$ (and $j_4$ for fifth-order tensors) are folded into a single index $j$. So $\bV_1^K\in J\times K$, where $J=J_1 \times J_2 \times J_3 (\times J_4)$.

Finally, to measure the quality of the reconstructed or enhanced database under study, we use the relative root mean square error (RRMSE), which is computed as
\begin{equation}
RRMSE=\sqrt{\frac{\sum_{k=1}^K||\textbf{v}_k-\textbf{v}^{approx.}_k||^2}{\sum_{k=1}^K||\textbf{v}_k||^2}},
\label{eq:rrmse}
\end{equation}
where $\textbf{v}_k$ and $\textbf{v}^{approx.}_k$ correspond to the real and approximated solutions, and $||\cdot||$ is the usual Euclidean norm.

\subsection{Singular value decomposition \label{sec:svd}}
Singular value decomposition (SVD) \cite{Sirovich87}, also known as proper orthogonal decomposition (POD)  \cite{Lumley}, is a method frequently applied for low-rank approximations and the extraction of meaningful information by eliminating redundant data and filtering noise from datasets. The method identifies the main patterns or coherent structures of the flow, representing the physics of the dynamical system under study \cite{VegaLeClaincheBook20}. It is important to note that, while POD and SVD are terms used interchangeably in the literature, SVD represents just one of two techniques capable of computing POD modes.

SVD algorithms factorizes the snapshot matrix $\bV_1^k$, eq. \eqref{ab0}, as follows
\begin{equation}
\bV_1^{K}\simeq\bW\,\bSigma\,\bT^\top,\label{eq:svd}
\end{equation}
where $(\cdot)^\top$ symbolizes the matrix transpose, $\bW$ is the matrix containing the POD modes (in columns), $\bT$ contains their associated temporal coefficients  and the diagonal of matrix $\bSigma$ contains the  singular values $\sigma_1,\cdots,\sigma_{K}$. 

The number of retained SVD modes $N$ is calculated based on a tunable threshold $\varepsilon_{svd}$ as
\begin{equation}
\sigma_{N+1}/\sigma_{1}\leq \varepsilon_{svd},\label{eq:TOLsvd}
\end{equation}
where  $W^\top W = T^\top T=$ are $N\times N-$unit matrix, with $N$ being the number of retained SVD modes, commonly referred to as \textit{spatial complexity}. It is important to distinguish between $J$, the {\it spatial dimension} of the dataset, and the spatial complexity $N$, where $N\leq min(J,K)$ \cite{LeClaincheVega17}.
The previous threshold, when applied, serves to reduce the dimensionality of the data from $J$ to $N$. When Singular Value Decomposition (SVD) is employed to clean experimental datasets from noise, the chosen threshold value typically aligns with the noise or uncertainty level inherent to the experiment. 

A variant of the SVD algorithm which has gained popularity in recent years  is high order SVD (HOSVD) \cite{DeLathawer,DeLathawer0}. 
%
%
HOSVD is used on datasets which are organized in tensor form, eq. \eqref{c43d}. The algorithm breaks down a tensor shaped datasets and applies SVD to each one of the tensor fibers. For example, HOSVD applied to a fifth-order tensor looks as so
\begin{equation}
V_{j_1j_2j_3j_4k}\simeq\sum_{p_1=1}^{P_1}\sum_{p_2=1}^{P_2}\sum_{p_3=1}^{P_3}\sum_{p_4=1}^{P_4}\sum_{n=1}^{N} S_{p_1p_2p_3p_4n}
   W^{(1)}_{j_1p_1} W^{(2)}_{j_2p_2} W^{(3)}_{j_3p_3} W^{(4)}_{j_4p_4}T_{kn}.  
   \label{eq:c10}
\end{equation}
$S_{p_1p_2p_3p_4n}$ is used to define what is known as the \emph{core tensor}, which is also a fifth-order tensor. The columns of the matrices $W^{(1)}$, $W^{(2)}$, $W^{(3)}$, $W^{(4)}$, and $T$ are commonly referred to as the \emph{decomposition modes}. The initial set of modes, in the proposed example, the columns of matrices $W^{(l)}$ for $l = $ to 1, 2, 3, and 4, represent the components and spatial variables that make up the dataset. These are known as the \emph{spatial HOSVD modes}. On the other hand, the columns of the matrix $T$ represent the modes associated with the time variable and are thus called the \emph{temporal HOSVD modes}.

The decomposition of singular values is now formed by five sets of values,
%
$\sigma^{(1)}_{p_1}$, 
$\sigma^{(2)}_{p_2}$, 
$\sigma^{(3)}_{p_3}$, 
$\sigma^{(4)}_{p_4}$, 
and $\sigma^t_{n}$,
%
which are arranged in decreasing order.
Similar to SVD, the HOSVD method (eq. \eqref{eq:c10}) is accurate without truncation. Nevertheless, truncation can be beneficial to filter noise, false artifacts, or reduce the data-dimensionality. Similar to SVD, the number of retained modes for each case often depends on a tunable threslhold as $\sigma_{P_1+1}/\sigma_{1}\leq \varepsilon_{svd_1}$, $\sigma_{P_2+1}/\sigma_{1}\leq \varepsilon_{svd_2}$, $\cdots$, $\sigma_{N+1}/\sigma_{1}\leq \varepsilon_{svd_5}$.

After applying truncation, HOSVD is re-written as 
\begin{equation}
 V_{j_1j_2j_3j_4k}\simeq\sum_{n=1}^{N} W_{j_1j_2j_3j_4n}\hat V_{kn},\label{eq:c15}
\end{equation}
where $W_{j_1j_2j_3j_4n}$ and $V_{kn}$ are the spatial and temporal modes, respectively, and $N$ is the spatial complexity which has been previously defined above. 

\subsection{Gappy SVD and gappy HOSVD\label{sec:gappy}}

The first application discussed in this article is the gappy SVD algorithm. This algorithm is designed to recover and repair datasets that have been compromised or have missing data. Such datasets typically arise from experiments involving sensor networks, where one or more sensors are malfunctioning, leading to the collection of corrupted data or even the absence of data altogether. In most instances, the conventional approach involves replacing the faulty sensors and rerunning the experiment. However, gappy SVD, also known as gappy POD, offers an alternative solution. It enables  to salvage datasets with corrupted or missing data by iteratively applying SVD to reconstruct the flawed or missing data, often referred to as \emph{gaps} in the original dataset. This data reconstruction is made possible through the decomposition properties of the SVD algorithm, which reorganizes the SVD modes based on their significance in reconstructing the original dataset.

\begin{figure}[H]
    \begin{center}
        \includegraphics[width=\textwidth]{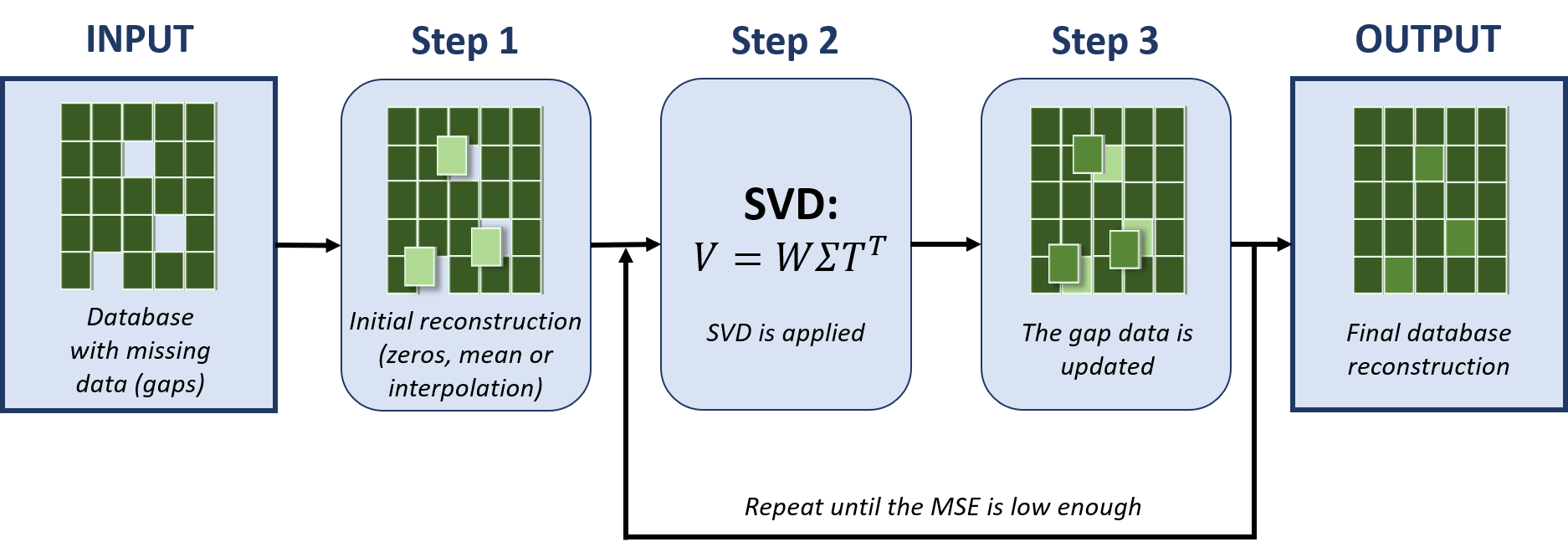}
        \vskip-0.1cm
    \end{center}
\vskip-0.5cm
\caption{Gappy SVD: sketch summarizing the methodology.\label{fig:gappy}}
\end{figure}

For two-dimensional databases, the gappy SVD algorithm uses the snapshot matrix from eq. \eqref{ab0}, with dimensions $J \times K$ as input data. For high dimensional datasets, the data is presented in tensor form (eq. \eqref{c43d}) and SVD is substituted with HOSVD. To simplify the mathematical explanation of the gappy SVD algorithm for both of the previously named factorization techniques, the method is explained particularizing the algorithm for a single snapshot of both a two-dimensional matrix form dataset and a three-dimensional tensor form dataset. 
%
To repair and reconstruct the dataset, gappy SVD applies the following steps. See more details of the algorithm in Refs. \cite{hetherington2023modelflows, Vega1HOSVDgappy, Venturietalgappy, Beckersetalgappy}.
\begin{itemize}
    \item[\bf Step 1.] Perform a first reconstruction to the original dataset, filling the initial gaps with zero, mean values or using linear or non-linear interpolation methods. This will give a new initial snapshot matrix. 
    \item[\bf Step 2.] Apply SVD to the previous matrix and reduce the previous matrix dimensions by retaining the singular values of $P'$, which can be tuned.  
    \item[\bf Step 3.] Reconstruct the new reduced snapshot matrix to update the initial gap values. 
    \item[\bf Step 4.] Calculate the mean square error of the gaps between iterations as
    \begin{equation}
        MSE_{gaps} = \dfrac{1}{N_{gaps}}\sqrt{\sum_{n=1}^{N_{gaps}} \lvert \widehat V^{i}-\widehat V^{i-1} \rvert},\label{eq:MSE}
    \end{equation}
    
    where $N_{gaps}$ is the number of gaps in the dataset. While $MSE_{gaps} > 10^{-6}$ (tunable), perform a new iteration 
    by repeating steps 2--4 on the new matrix calculated in Step 3. This will give the final reconstructed matrix.
\end{itemize}
%
Figure \ref{fig:gappy} illustrates the previously described steps in the form a flowchart.


\subsection{SVD superresolution algorithm \label{sec:superres}}
SVD for resolution enhancement, also known as the \textit{SVD superresolution algorithm}, is similar to gappy SVD in the sense that the algorithm also applies SVD iteratively to an input dataset, enhancing its resolution thanks to the properties of the SVD algorithm which, as previously mentioned, re-organized the SVD modes based on their contribution when reconstructing the original dataset. To explain the methodology, the algorithm is also particularized for an individual snapshot $\bv_k$, which is organized in matrix form $\bV^{DS}$. 

The algorithm is summarized as follows. More details can be found in \cite{hetherington2023modelflows}.
\begin{itemize}
    \item[\bf Step 1.] Set the dimensions $\widehat N_1\times\widehat N_2 = J$ for the enhanced resolution dataset. A $NaN$ base matrix is created with said dimensions and uniformly distribute the initial under-resolved (down-sampled) data.
    
    \item[\bf Step 2.] Perform a first reconstruction to the base matrix, assigning an initial value to the $NaN$ data (referred to as new data $n$) points using linear or non-linear interpolation between the previously distributed down-sampled matrix data. The interpolation is created by using Qhull to triangulate the input data \cite{barber1996quickhull}, and then applying linear barycentric interpolation \cite{hormann2014barycentric} within each triangle.
    
    \item[\bf Step 3.] Apply SVD to the previous matrix, and set (tune) the number of singular values to retain. 
    
    \item[\bf Step 4.] Reconstruct the reduced matrix. 
    
    \item[\bf Step 5.] Calculate the mean square error of the values between iterations as 
    \begin{equation}
        MSE_{n} = \dfrac{1}{M_{n}}\sqrt{\sum_{m=1}^{M_{n}} \lvert \widehat V^{i}- \widehat V^{i-1} \rvert},
    \end{equation}
    where ${}_n$ refers to the new data, $M_{n}$ is the amount of new data created in the dataset in order to match the desired dimensions. While $MSE_{n} > 10^{-6}$, perform a new iteration by repeating steps 3--5 on the new matrix calculated in Step 4. The final matrix is the reconstructed matrix with enhanced resolution. 
\end{itemize} 

\begin{figure}[H]
    \begin{center}
        \includegraphics[width=\textwidth]{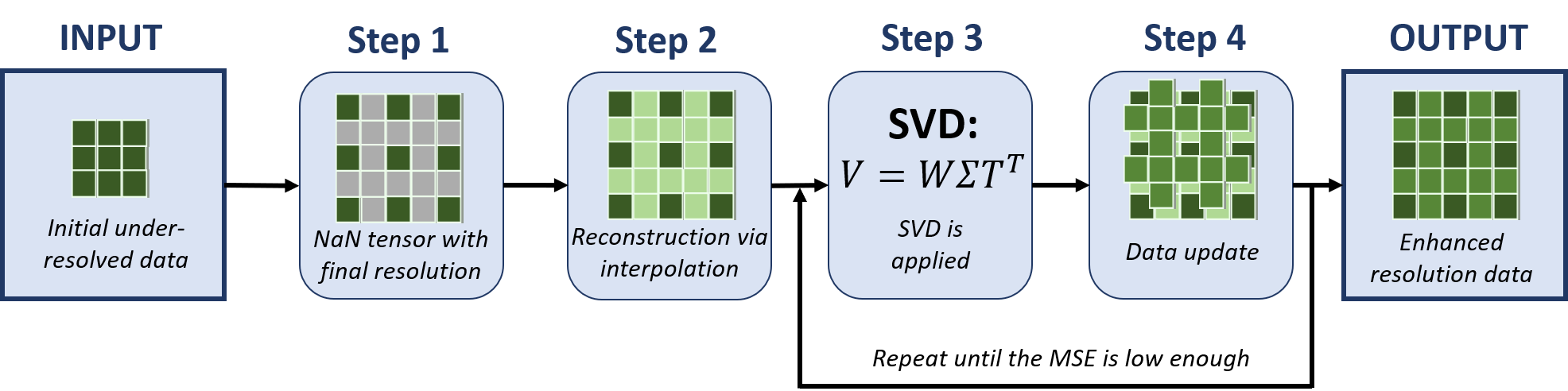}
        \vskip-0.1cm
    \end{center}
\vskip-0.5cm
\caption{SVD superresolution algorithm: sketch summarizing the methodology.\label{fig:svdsuper}}
\end{figure}

The previous steps are illustrated in Fig. \ref{fig:svdsuper}.


For three-dimensional or high dimensional data, the same preceding steps are followed, but the SVD algorithm is replaced by the HOSVD. Moreover, HOSVD is employed to enhance the resolution of databases that involve temporal components. The temporal information (which can be adjusted) may either remain consistent throughout the algorithm, thereby improving only the spatial components' resolution, or it can be expanded to interpolate time instances that were not originally present in the dataset.
See more details regarding this algorithm and some applications in Refs. \cite{VegaSVDResolution,NoureletalNMR}.

\subsection{Deep learning superresolution (DL superresolution)\label{sec:dlsuperres}}
\begin{figure}[H]
	\centering
    \includegraphics[width=\textwidth]{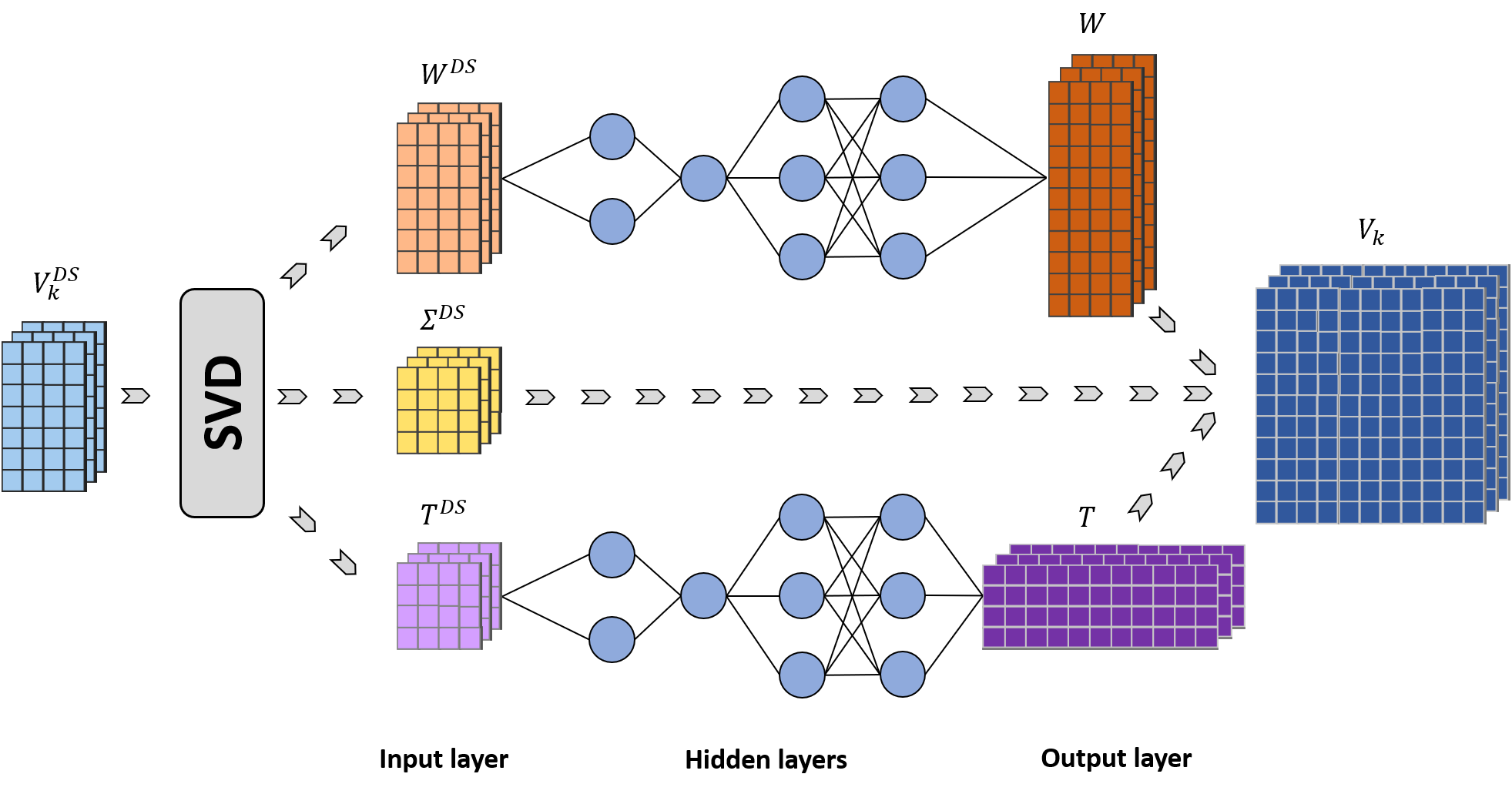}
	\caption{Reconstruction of databases combining SVD and deep learning architectures. Sketch of the methodology. \label{fig:NN_arq}}
\end{figure}

Dataset resolution enlargement can also be achieved by using hybrid ROMs, which combine modal decomposition algorithms with different neural network architectures. Hybrid ROMs can be defined in two steps. Figure \ref{fig:NN_arq} presents an example sketch of this hybrid ROM, formed by 5 layers. 

The first step of the algorithm consists in applying SVD to the snapshot matrix as follows:

\begin{equation}
\bV_1^{K}\simeq\bW\,\bSigma\,\bT^\top,\label{ab20v2}
\end{equation}

In this case, the snapshot matrix $\bV_1^K$ eq. \eqref{ab0} is renamed as $\bV^{DS}$, as it represents a down-sampled dataset which is formed by a reduced number of data points that contain the temporal measurements of a velocity or pressure field, or other variables, which depend on the data that is being studied. For each snapshot, the down-sampled snapshot matrix $\bV^{DS}$ has dimensions $N_1\times N_2$, with $N_1 \times N_2 <J$. For two-dimensional datasets, $N_1$ and $N_2$ correspond to the streamwise and normal components, respectively, while for three-dimensional databases, $N_2$ contains both the normal and spanwise components (although the three spatial components can be re-organized differently into dimensions $N_1$ and $N_2$ according to the problems requirements). Particularizing for a single snapshot, eq. \eqref{ab20v2} is applied, and the data is re-organized into a down-sampled matrix, which is re-written as 
\begin{equation}
\bV^{DS}\simeq\bW^{DS}\,\bSigma^{DS}\,(\bT^{DS})^\top.\label{ab20v3},
\end{equation}

where the diagonal of matrix $\bSigma^{DS}$, whose dimensions are $P' \times P'$, contains the retained singular values. The remaining matrices are $\bW^{DS}$ and $\bT^{DS}$, which have dimensions $N_1 \times P'$ and $N_2 \times P'$, respectively. 

In step two, the process involves feeding these last two matrices into a deep learning model, which in turn produces the reconstructed dataset. This model architecture consists of two decoders, each dedicated to one of the matrices, and they operate in parallel. These decoders expand the dimensions from $N_1$ and $N_2$ of the down-sampled matrices to match $\widehat N_1$ and $\widehat N_2$, where $\widehat N_1 \times \widehat N_2 = J$. The resulting matrices are denoted as $\bW$, with dimensions $N_1 \times P'$, and $\bT$ with dimensions $N_2 \times P'$.

The outputs of both decoders, along with $\bSigma^{DS}$, are then forwarded to the output layer. This output layer generates the reconstructed solution for each snapshot $K$, with a spatial dimension of $J$, in the following manner: 

\begin{equation}
\bV_k\simeq \bW\,\bSigma^{DS}\,(\bT)^\top,\label{ab20v4}
\end{equation}

where $\bV_k$, with dimensions $\widehat N_1 \times \widehat N_2$, is the enhanced spatial resolution for a singular snapshot $K$. Note that this architecture makes it possible to also enhance the temporal resolution of any dataset by using the same methodology to calculate new snapshots between the preexisting ones. More details can be found in \cite{diaz2024deep}.
The RRMSE from eq. \eqref{eq:rrmse} is used to calibrate the weights and bias of the model by comparing the reconstructed dataset with the original solution. 

\begin{figure}[H]
	\centering
	\includegraphics[width=0.8\columnwidth]{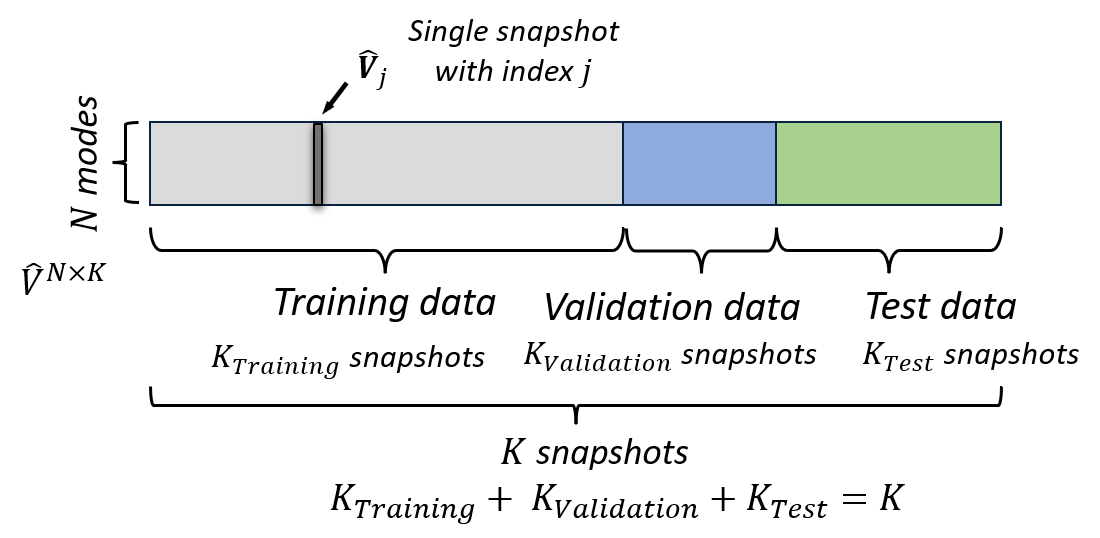}  
    \caption{Sketch with the training, validation and test set distribution for the deep learning module. \label{fig:ML2}}
\end{figure}

It is important to understand that, in order to train any hybrid ROM, or neural network in general, the model must be trained using preexisting data. The dataset must be split into three parts: training, validation and test data. This split is illustrated in Fig. \ref{fig:ML2}. 

This way, the model will learn the necessary patterns to perform data resolution enhancement using the training data, and the results are compared to those obtained using the validation data. Once the model training is complete, then the test data is forwarded to the model to test its performance on data that its has not previously been exposed to. This model can also use this low resolution data to make high resolution temporal predictions, generating new data.

\subsection{Error analysis using uncertainty quantification \label{sec:uq}}
Uncertainty quantification is the error analysis technique used to estimate the reconstruction error using statistics. This is, the error between the original dataset and the reconstruction solution provided by the previously described methods.

Consider a three-dimensional dataset, organized into a fifth-order tensor, following eq. \eqref{c43d}. The reconstruction error $\epsilon$ for each component ($j_1 = 1, 2, 3$ for $J_1 = 3$) is computed as the difference between $\bV_{j1}$ and $\bV_{j1}^{rec}$. Therefore, each component reconstruction error is calculated as follows:
\begin{equation}
\begin{split}
\epsilon_{u} = V_{1J_2J_3J_4J_K} - V^{rec}_{1J_2J_3J_4J_K}, \\ \epsilon_{v} = V_{2J_2J_3J_4J_K} - V^{rec}_{2J_2J_3J_4J_K}, \\ 
\epsilon_{w} = V_{3J_2J_3J_4J_K} - V^{rec}_{3J_2J_3J_4J_K},\label{eq:epsUV} 
\end{split}
\end{equation}
where $\epsilon_{u}$, $\epsilon_{v}$, $\epsilon_{w}$ are fourth-order tensors that share the same dimensions, $J_2 \times J_3 \times J_4 \times K$. In fluid mechanics problems, these reconstruction errors are then normalized based on the maximum absolute error value of each velocity component as
\begin{equation}
\frac{\epsilon_{u}}{|\epsilon_{u, max}|}, \quad \frac{\epsilon_{v}}{|\epsilon_{v, max}|}, \quad \frac{\epsilon_{w}}{|\epsilon_{w, max}|},\label{eq:epsnormUV} 
\end{equation}

where $\epsilon_{u, max}$, $\epsilon_{v, max}$ and $\epsilon_{w, max}$ represent the highest reconstruction error value for each component, and $|\cdot|$ denotes the absolute value. The reconstruction errors are represented using two different plots. The first plot displays the probability density function of the reconstruction error for each component. This plot serves the purpose of examining the probability of each error value. The probability curves tend to follow a normal distribution centered in zero. When these curves are narrow and tall, it signifies low uncertainty, meaning that a given error value is most likely to be zero, meaning the reconstruction is accurate. Conversely, when the curves wide and flat, it suggests high uncertainty. In this scenario, the reconstruction error for a given data point could potentially take on a wide range of values, which are displayed on the horizontal axis. 

The second plot consists of an absolute error contour map, and is used to perform a visual analysis of the reconstruction error. Given a singular snapshot of $|\epsilon|$, when its contour map is traced, it is possible to visually interpret in which areas the reconstruction error is higher or lower.  

\section{Test cases\label{sec:database}}
This section presents the test cases under study, which are a three-dimensional numerical simulation of a cylinder, a turbulent experimental cylinder flow, and an experimental rough plate zero pressure gradient turbulent boundary layer flow. These test cases have been specifically selected to test the presented methods reconstruction performance when applied to high dimensional numerical simulation data and, specially, experimental data which contains measurements with noise.

All test cases consist of fluid dynamics experiments or numerical simulations which are governed by the Navier-Stokes equations. These equations for a viscous, incompressible and Newtonian flow are:

\begin{equation}
\vec{\bV} \cdot \nabla = 0
\end{equation}

\begin{equation}
\frac{\partial u}{\partial t} + (\vec{\bV} \cdot \nabla) \vec{\bV} = -\nabla p + \frac{1}{Re} \Delta \vec{\bV},
\end{equation}

where $\vec{\bV}$ is the velocity vector, $p$ describes the pressure, and Re is the Reynolds number, which is defined as Re $= \rho U L / \mu$, where $\rho$ is the fluid density and $\mu$ is the dynamic viscosity. These equations are non-dimensionalized using the characteristic length $L$ and velocity $U$.




\subsection{Three-dimensional circular cylinder \label{Cyl3D}}
The first dataset discussed in this context is a numerical simulation of laminar flow around a three-dimensional circular cylinder. Such datasets are frequently selected as standard benchmarks for evaluating and testing new algorithms and models. 

The behavior of the cylinder is closely linked to the concept of the Reynolds number, which is defined based on the cylinder diameter, denoted as $D$. When the Reynolds number is low, the flow remains steady. However, as it approaches Re $\approx 46$, a Hofp bifurcation occurs, leading to an unsteady flow characterized by a von Karman vortex street \cite{jackson1987finite}. These oscillations maintain a two-dimensional nature until Re $\approx 189$, at which point a second bifurcation takes place, causing the flow to transition into a three-dimensional state with specific wavelengths in the spanwise direction \cite{barkley1996three}. 
At high Reynolds values, the flow becomes fully turbulent.

The flow of the selected cylinder dataset has a Reynolds number of Re $= 280$ \cite{LeClaincheetalFDR18}, which is based on the cylinder diameter $D$ and the free flowing fluid velocity $U_{\infty}$. The numerical simulations are performed using the open source solver Nek5000 to solve incompressible Navier-Stokes equations. This solver uses spectral elements methods as spatial discretization. The database used for the present analysis is described in \cite{VegaLeClaincheBook20}. 

\begin{figure}[H]
	\begin{center}
    \includegraphics[width=\textwidth]{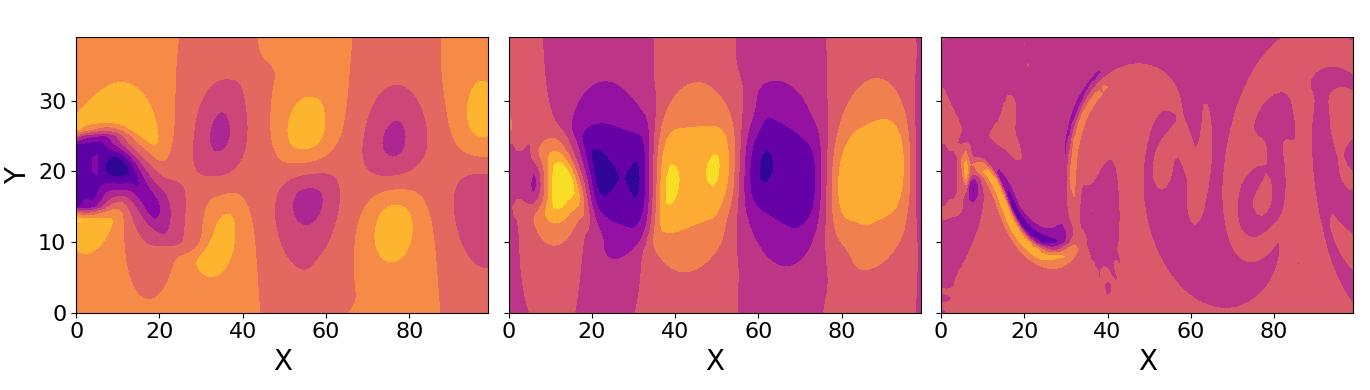}
\vskip-0.1cm
\end{center}
\vskip-0.5cm
	\caption{Streamwise (left), normal (middle) and spanwise (right) velocities of a representative snapshot of the three-dimensional cylinder dataset at Re $ = 280$ from \cite{VegaLeClaincheBook20}. \label{fig:3dcyl}}
\end{figure}

The dataset consists of 599 snapshots equidistant in time, with time step $\Delta t$ = $1$ $ms$. Only the last 299 snapshots of this dataset are used to represent the saturated flow regime (the previous one represent the transient regime of the simulations). The dataset consists of $N_{x}$ = 100 points in the streamwise direction, $N_{y}$ = 40 in the normal direction, and $N_{z}$ = 64 in the spanwise direction. The flow field velocity components are defined by $U$ for the streamwise velocity, $V$ for the normal velocity, and $W$ for the spanwise velocity, so the number of components $N_{comp} = 3$, and their velocity contours are presented in Fig. \ref{fig:3dcyl}.

\subsection{Turbulent circular cylinder \label{VKI4000}}
This first experimental dataset, which belongs \cite{mendez2020multiscale}, and consists of a turbulent flow passing a cylinder in steady and transient conditions, producing Von Karman wake vortices of with variant frequency. The experiment was developed in the L10 low-speed wind tunnel of the von Karman Institute, equipped with a TR-PIV system.

The cylinder has a diameter of $D = 5$ $mm$ and is $L = 200$ $mm$ in length. The experiment was conducted in transient conditions with variant free stream velocity. The full database consists of approximately $N_t = 13500$ snapshots which elapses during 4.5 seconds.

\begin{figure}[H]
	\begin{center}
    \includegraphics[width=\textwidth]{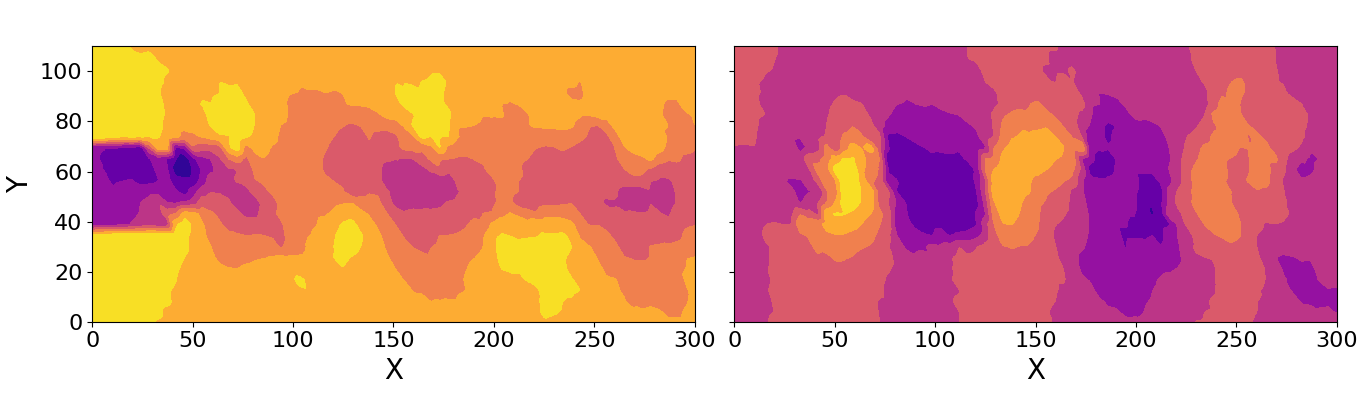}
\vskip-0.1cm
\end{center}
\vskip-0.5cm
	\caption{Streamwise (left) and normal (right) velocities of a representative snapshot of the turbulent Re $ = 4000$ cylinder dataset from \cite{mendez2020multiscale}. \label{fig:vki4000}}
\end{figure}

The dataset contains three different states: two stationary states, obtained at Re $  \approx 4000$ and Re $  = 2600$, and the transitory state between both of these. 
In this article the first steady state is used, which corresponds to the Re $  \approx 4000$ flow. This dataset consists of $N_x = 301$ points in the streamwise direction and $N_y = 111$ in the normal direction. The two velocity components ($N_{comp} = 2$), streamwise $U$ and normal velocity $V$, have been registered during $N_t = 4000$ snapshots with a time step of $\Delta t = 0.33$ $ms$. Figure \ref{fig:vki4000} presents the first snapshot of both velocity components of this dataset.

\subsection{Rough flat plate turbulent boundary layer \label{BL}}
The last dataset that has been selected is also experimental, and belongs to \cite{lawrence2021piv}. The experiment was conducted in the great horizontal water tunnel located in the Laboratory for Turbulence Research in Aerospace and Combustion (LTRAC), at University of Monash (Australia). 


The rough flat plate used in this experiment has a relative roughness height equivalent to $k/\delta_e = 0.052$, with initial roughness height $\lambda_x = 4$, located $k_0 = 0.34$ $mm$ from the origin. The roughness height growth rate is $\frac{d_k}{d_x} = 0.00139$. 

The velocity fields were acquired in the zero-pressure-gradient turbulent boundary layer, the shear layer, and near and far wakes. A high velocity system was used to obtain sufficiently time-resolved data which was necessary to obtain the flow spectral composition information. 

The above mentioned velocity fields were obtained performing experimental two-component two-dimensional particle image velocimetry (PIV 2C-2D). Different PIV resolution and velocity systems were used for each flow region. A high velocity and high resolution PIV system was used in the turbulent boundary layer and the near wake regions, high speed PIV was used for the shear layer region, and high resolution PIV for the far wake region.

\begin{figure}[H]
	\begin{center}
    \includegraphics[width=\textwidth]{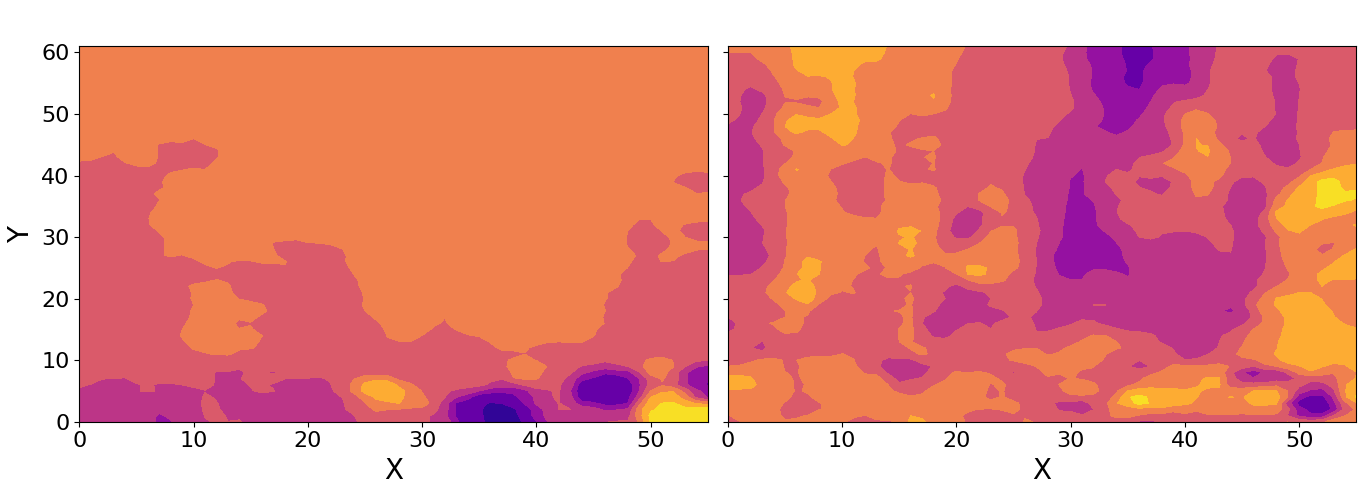}
\vskip-0.1cm
\end{center}
\vskip-0.5cm
	\caption{Streamwise (left) and normal (right) velocities of a representative snapshot of the turbulent boundary layer dataset from \cite{lawrence2021piv}.  \label{fig:tbl}}
\end{figure}

The dataset consists of two velocity components ($N_{comp} = 2$), streamwise $U$ and normal $V$ (see Fig. \ref{fig:tbl}), in the proximity of the boundary layer formed on rough surface flat plate.

The dataset is composed by $N_x = 56$ vectors in the streamwise direction and $N_y = 62$ vectors in the wall-normal direction. Both of the velocity components are measured during $N_t = 6297$ snapshots, with a time step equal to  $\Delta t = 4$ $ms$.

\section{Results\label{sec:results}}
This section presents the results obtained applying the algorithms from Sec. \ref{sec:methodology} to the test cases described in the previous section. First, the reconstruction results of the gappy SVD algorithm when applied on modified versions of the test cases are presented. These modified version contain different amounts of randomly distributed missing data (gaps) to be reconstructed, based on three number of retained SVD modes cases, which are consistent and robust throughout the results section. Next, the SVD superresolution algorithm results are demonstrated by applying this method to under-resolved versions of each test case to augment their resolution to that of the test case. Also, the deep learning superresolution model is applied to these same datasets, and the resolution enhancement results obtained using both methods are then compared. The reconstruction error metrics used for all of these methods are $RRMSE$ (eq. \eqref{eq:rrmse}) and uncertainty quantification (Sec. \ref{sec:uq}).

\subsection{Data repairing with gappy SVD \label{sec:gappyresults}}

The gappy SVD algorithm has been applied on modified versions of the test cases presented in Sec. \ref{sec:database}, generating gaps by randomly removing data from these datasets. Three variant datasets have been created for each case by substituting 20\%, 40\% and 60\% of the data with $NaN$ values to create gaps to repair. Since, as explained in Sec. \ref{sec:gappy}, the gappy SVD algorithm uses SVD (or HOSVD) to repair the data, three values for retained SVD modes have been selected to test the algorithms reconstruction capabilities. These values are 5, 10 and 15, which have been selected to test the method reconstruction capabilities and limitations when decreasing the number of retained modes and, therefore, the amount of representative information used for the reconstruction. Also, two cases of initial reconstruction have been selected to fill the gaps, these being linear interpolation and zeros.

\begin{figure}[H]
	\begin{center}
    \includegraphics[width=\textwidth]{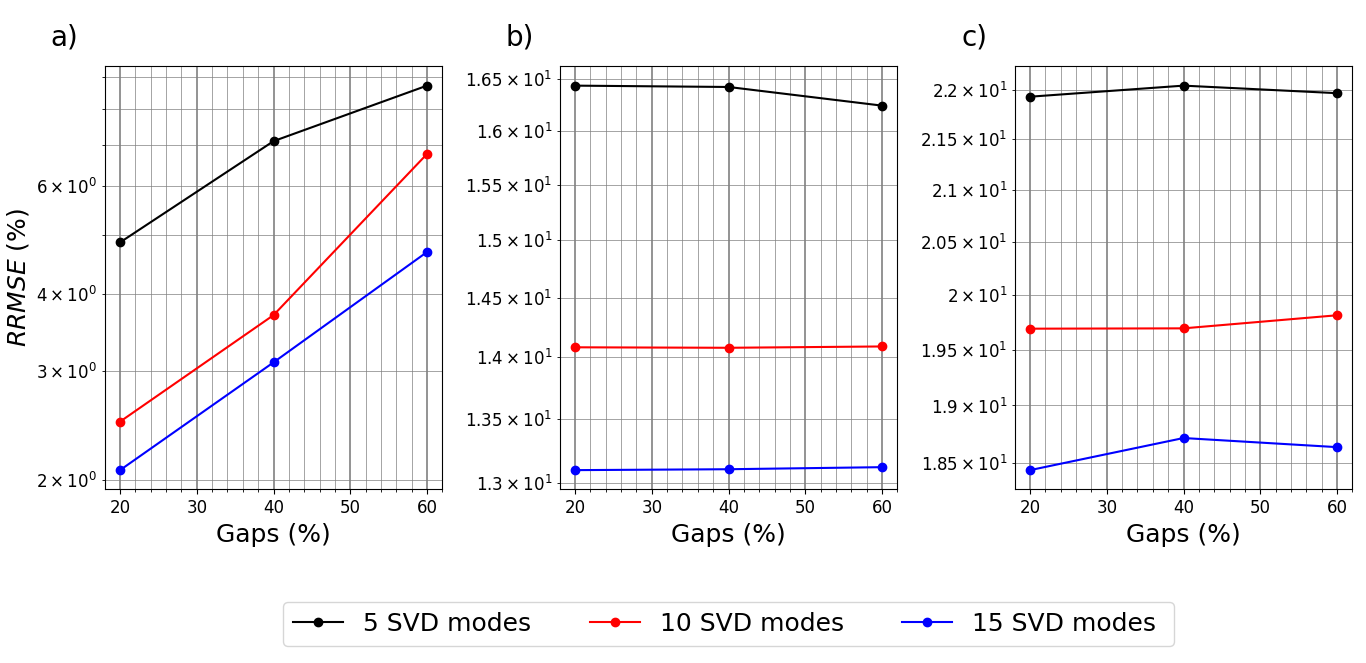}
\vskip-0.1cm
\end{center}
\vskip-0.5cm
	\caption{Reconstruction $RRMSE$ evolution for each case, these being 20\%, 40\% and 60\% of gaps, and 5, 10 and 15 retained singular values. The initial reconstruction has been performed using linear interpolation: a) presents the results for the three-dimensional cylinder at Re $= 280$, b) the turbulent Re $= 4000$ cylinder, and c) the turbulent boundary layer. \label{fig:rrmseevolgappy}}
\end{figure}

Table \ref{tab:rrmsegappy} gathers all $RRMSE$ scores, which allow to gauge how well gappy SVD has rebuilt the datasets when comparing the reconstruction results to the ground truth for all case combinations (gaps and SVD modes). Figure \ref{fig:rrmseevolgappy} accompanies the previous table and serves to illustrate these same results. Note that these results have been obtained by applying linear interpolation to perform the initial data reconstruction.

\begin{table}[H]
        \centering
        \begin{tabular}{|l  l  l  l l|}
        \hline
        \rowcolor{Gray}
        \hline
        \textbf{Datasets} & \textbf{modes} & \textbf{20\% gaps} & \textbf{40\% gaps} & \textbf{60\% gaps}
        \\ \hline \hline
            Cyl3D Re280 & 5 & 4.856 & 7.089 & 8.724\\
            & 10 & 2.485 & 3.7 & 6.757\\
            & 15 & 2.073 & 3.101 & 4.684\\
            \hline
            Cyl2D Re4000  & 5 & 16.437  & 16.424 & 16.244\\
            & 10 & 14.082 & 14.078 & 14.089\\
            & 15 & 13.096 & 13.103 & 13.119\\
            \hline
            TBL  & 5 & 21.93 & 22.043 & 21.966 \\
            & 10 & 19.69 & 19.693 & 19.813 \\
            & 15 & 18.438 & 18.715 & 18.636 \\
            \hline
        \end{tabular}
        \caption{Reconstruction $RRMSE$ (\%) values when repairing the test case dataset variants with 20\%, 40\% and 60\% of gaps by performing an initial reconstruction using linear interpolation and using gappy SVD with 5, 10 and 15 retained SVD modes. \textit{Cyl3D Re280} represents the three-dimensional cylinder at Re $= 280$, \textit{Cyl2D Re4000} is the turbulent Re $= 4000$ cylinder, while \textit{TBL} symbolizes the turbulent boundary layer dataset.}
        \label{tab:rrmsegappy}
    \end{table}

For all test cases, the reconstruction $RRMSE$ presents a tendency to decrease when the number of gaps is low, this being due to less data requiring reconstruction. The error also decreases when the number of SVD modes increases since more of the most relevant information that describes the dataset is made available for the reconstruction, providing a more accurate and detailed solution. When increasing the number of gaps, the reconstruction $RRMSE$ decay is more notorious for the three-dimensional Re $= 280$ dataset than the rest of test cases. This is because this dataset is laminar, so it contains less noise, consequently, it is easier to reconstruct. In addition, this dataset contains more data points, so the number of gaps is higher. On the other hand, the turbulent Re $= 4000$ cylinder and turbulent boundary layer are both turbulent experimental datasets that contain vast amounts of noise, and when the data is reconstructed after HOSVD is applied using a low number of retained modes, the noise is filtered out. Therefore, when comparing the original dataset with the reconstruction, the $RRMSE$ is higher, since the noise has not been reconstructed.

\begin{figure}[H]
	\begin{center}
    \includegraphics[width=\textwidth]{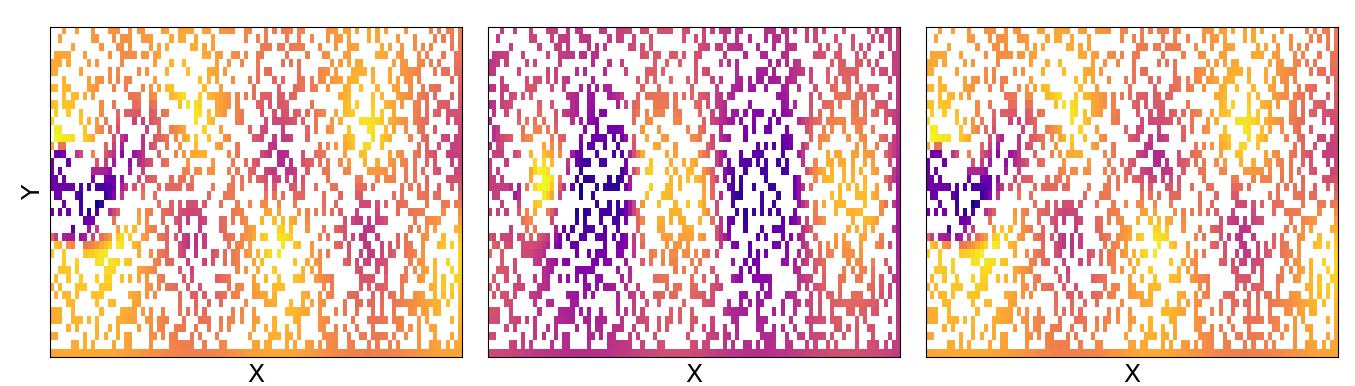}
\vskip-0.1cm
\end{center}
\vskip-0.5cm
	\caption{Streamwise (left), normal (middle) and spanwise (right) velocity components of a representative snapshot of the dataset with 60\% of gaps of the three-dimensional cylinder at Re $= 280$. \label{fig:gappyinitialcyl3d}}
\end{figure}

The dataset with 60\% of gaps of the three-dimensional cylinder at Re $= 280$ can be seen in Fig \ref{fig:gappyinitialcyl3d}.

\begin{figure}[H]
	\begin{center}
    \includegraphics[width=\textwidth]{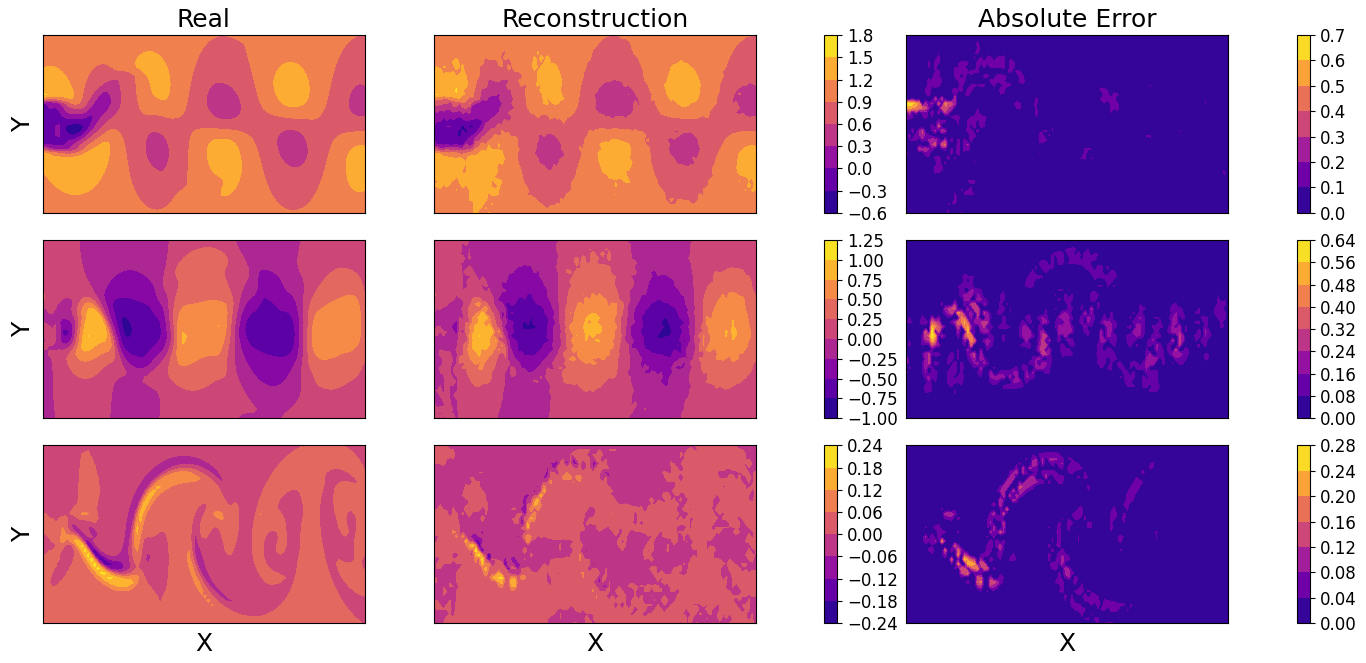}
\vskip-0.1cm
\end{center}
\vskip-0.5cm
	\caption{Gappy SVD reconstruction results for the dataset with 60\% of gaps of the three-dimensional cylinder at Re $= 280$ using 5 retained SVD modes. From left to right and top to bottom: the real data, the gappy SVD reconstruction and the absolute error of the streamwise, normal, and spanwise velocities, for the highest absolute error snapshot of each velocity component. \label{fig:gappycyl3d_60_5}}
\end{figure}

The reparation results for the highest reconstruction $RRMSE$ snapshot of each velocity component of the previous dataset, for 5 retained SVD modes, can be seen in Fig. \ref{fig:gappycyl3d_60_5}. This figure compares the original data to the reconstruction solution, and also includes the absolute error contour map, which allows to visualize where the highest reconstruction error is located for each component (as expected, is in the near field wake of the cylinder, where the most complex dynamics occurs).

Despite more than half of the dataset missing data, gappy SVD is capable of reconstructing the dataset with reconstruction $RRMSE = 8.724$, as seen in Tab. \ref{tab:rrmsegappy}.  The figure shows that the absolute error values scale is in the order of $10^{-1}$, with peak error points being minimum and sparse in comparison to those where the error is 0. Also, the highest absolute error for the streamwise and normal velocities is half the maximum velocity of the original data. For the spanwise velocity, notice that the velocity and absolute error maximum values are practically the same. This is due to the spanwise velocity being low, matching the error tolerance scale. 

\begin{figure}[H]
	\begin{center}
    \includegraphics[width=\textwidth]{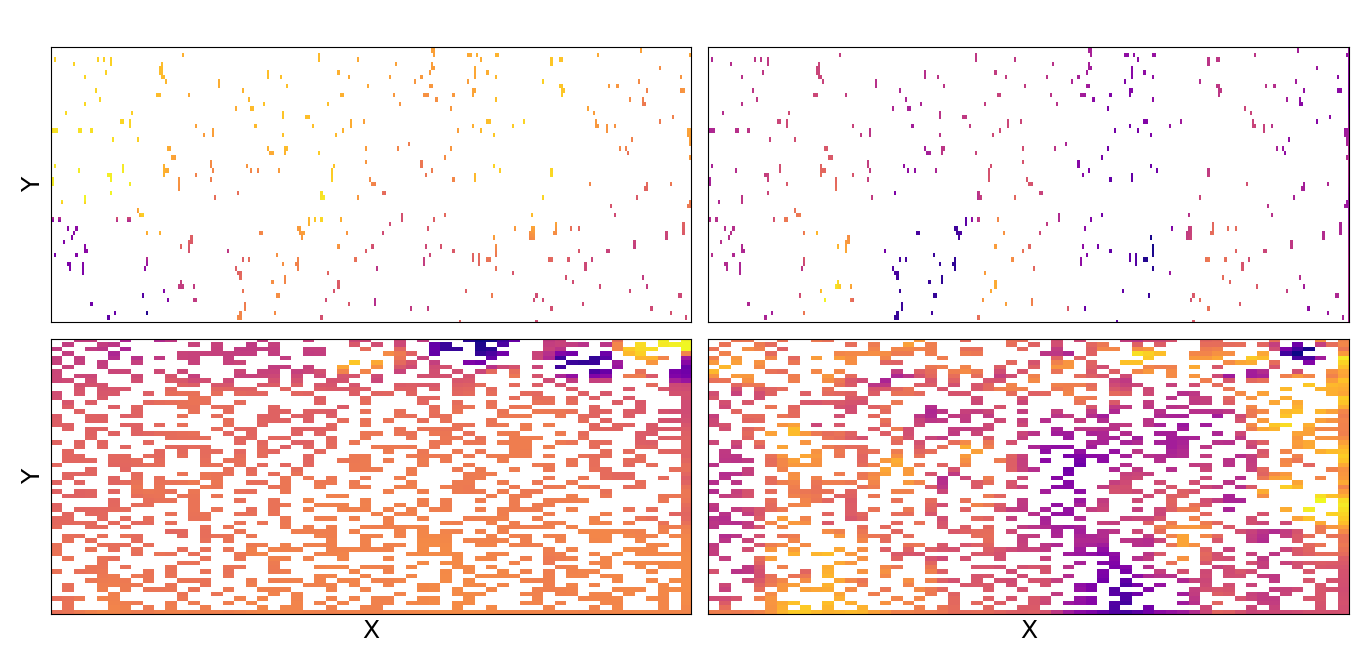}
\vskip-0.1cm
\end{center}
\vskip-0.5cm
	\caption{From top to bottom: streamwise (left) and normal (right) velocity components of a representative snapshot of the datasets with 60\% of gaps of the turbulent Re $= 4000$ cylinder and the turbulent boundary layer. \label{fig:gappyvkitbl}}
\end{figure}

Figure \ref{fig:gappyvkitbl} shows the two velocity components analyzed of the turbulent Re $= 4000$ cylinder and the turbulent boundary layer datasets with 60\% gaps (most complex cases). 

\begin{figure}[H]
	\begin{center}
    \includegraphics[width=\textwidth]{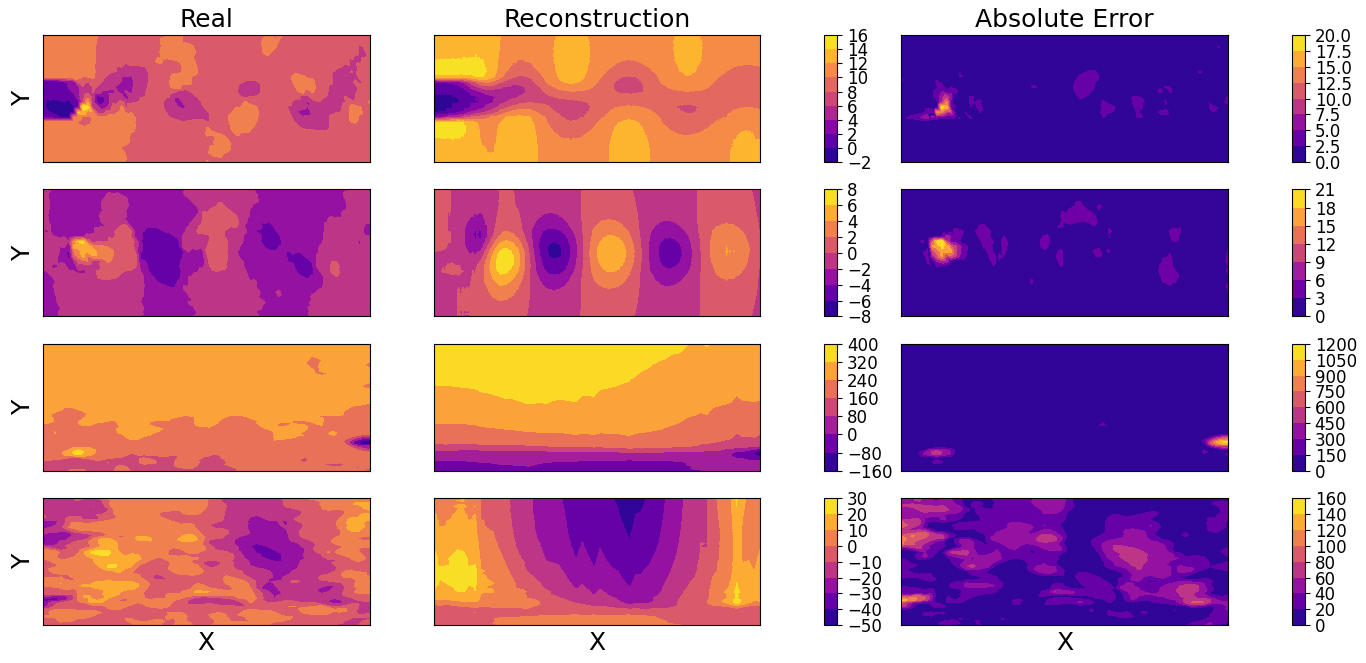}
\vskip-0.1cm
\end{center}
\vskip-0.5cm
	\caption{Gappy SVD reconstruction results for 60\% gaps and 5 SVD modes. From left to right and top to bottom: the real data, gappy SVD reconstruction and absolute error for the highest error snapshot of the streamwise and normal velocities of the turbulent Re $= 4000$ cylinder, shown in the first two rows, and the streamwise and normal velocities of the turbulent boundary layer, shown in the last two rows. \label{gappygroupedgappyvkitbl_60_5}}
\end{figure}

Figure \ref{gappygroupedgappyvkitbl_60_5} presents the reconstruction results when applying gappy SVD to the previous datasets. Note that the combination of 5 SVD modes and 60\% gaps is presented for all test cases since it provides the highest reconstruction error. This is used to prove the gappy SVD algorithm reconstruction capabilities in the most disadvantageous circumstances.

Similar to the three-dimensional cylinder at Re $= 280$ results presented in Fig. \ref{fig:gappycyl3d_60_5}, the majority of absolute error values for both test case results are 0, while only a small sparse amount of peak error values can be seen. Also, these peak absolute error values are the same scale as the velocity, which explains why the reconstruction $RRMSE$ displayed in Tab. \ref{tab:rrmsegappy} is much higher for both of these test cases in comparison to the three-dimensional cylinder at Re $= 280$ case. In other words, despite the majority of absolute error values for all three test cases with 60\% gaps being 0, the $RRMSE$ score for the  turbulent cylinder and turbulent boundary layer are higher since the peak error values also follow this tendency. The cause of this is the quantity of noise from the experimental database, which does not allow to properly calculate the $RRMSE$ of the databses, as explained at the beginning of this subsection. This fact is also reflected When comparing the original and reconstruction columns from Fig.\ref{fig:gappyvkitbl}: notice that gappy SVD eliminates all noise from the data, this being the reason why, when comparing the original data with the reconstruction, the $RRMSE$ and absolute error are both higher since the reconstruction does not contain the noise that the original data has.

\begin{figure}[H]
	\begin{center}
    \includegraphics[width=\textwidth]{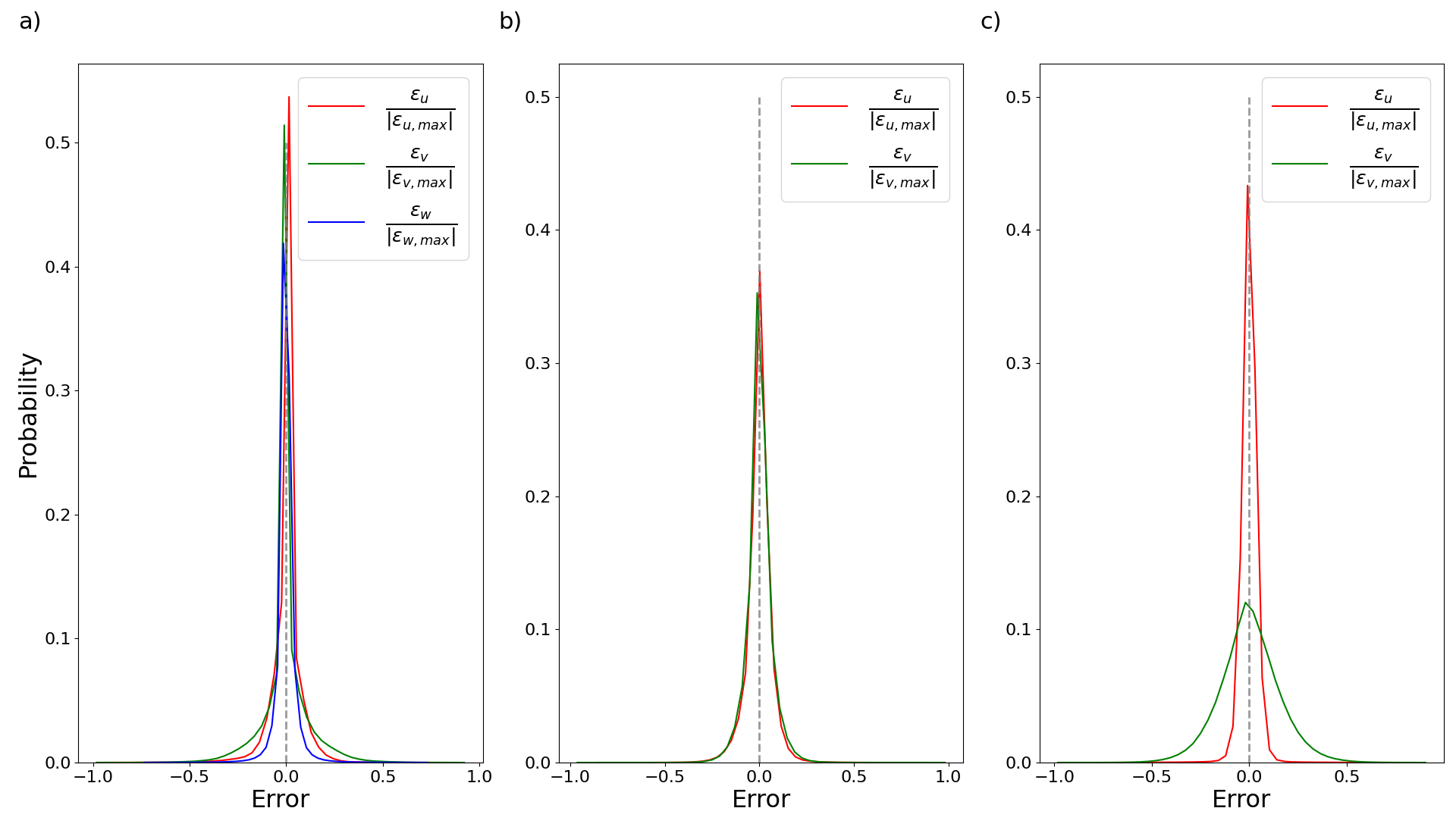}
\vskip-0.1cm
\end{center}
\vskip-0.5cm
	\caption{Gappy SVD reconstruction error probability density functions for each velocity component of: a) the three-dimensional cylinder at Re $= 280$, b) turbulent Re $= 4000$ cylinder, and c) turbulent boundary layer dataset, all for 60\% gaps and 5 retained SVD modes. \label{fig:pdfgappy}}
\end{figure}

The reconstruction error probability density functions for all velocity components of the three test cases can be seen in Fig. \ref{fig:pdfgappy}. 

The reconstruction error probability distribution functions for all three velocity components of the three-dimensional cylinder at Re $= 280$ (Fig. \hyperref[fig:pdfgappy]{\ref*{fig:pdfgappy}(a)}) apparently follow similar distributions, but are slightly offset from zero, meaning that the error values with higher probability are very close to zero, but not exactly this value, and, therefore, these do not follow a normal distribution. The streamwise velocity reconstruction error probability density function is faintly left skewed, while the normal and spanwise distributions are gently right skewed. This is due to the scale of the reconstruction error values being low, meaning there is more precision when distinguishing different error values on a smaller scale. On the other hand, the probability density functions for the streamwise and normal velocity reconstruction errors of the turbulent Re $= 4000$ cylinder (Fig. \hyperref[fig:pdfgappy]{\ref*{fig:pdfgappy}(b)}) and turbulent boundary layer (Fig. \hyperref[fig:pdfgappy]{\ref*{fig:pdfgappy}(c)}) all follow a normal distribution. Notice that both distributions for the turbulent Re $= 4000$ cylinder are practically identical, while the distributions of the turbulent boundary layer differ, with the normal velocity reconstruction error probability distribution function being much wider, meaning that there is a higher level of uncertainty contained in the reconstruction of this component.

As mentioned in Sec. \ref{sec:gappy}, in the first step of the gappy SVD algorithm, an initial reconstruction can also be performed by substituting all $NaN$ values with zero. Table \ref{tab:gappyzeros} shows the reconstruction $RRMSE$ scores for all the proposed cases when applying gappy SVD with the initial reconstruction completed with zeros.

\begin{table}[H]
        \centering
        \begin{tabular}{|l  l  l  l l|}
        \hline
        \rowcolor{Gray}
        \hline
        \textbf{Datasets} & \textbf{modes} & \textbf{20\% gaps} & \textbf{40\% gaps} & \textbf{60\% gaps}
        \\ \hline \hline
            Cyl3D Re280 & 5 & 10.721 & 12.574 & 16.176\\
            & 10 & 8.69 & 12.574 & 16.176\\
            & 15 & 10.095 & 12.574 & 23.472\\
            \hline
            Cyl2D Re4000  & 5 & 16.437  & 16.241 & 17.675\\
            & 10 & 14.082 & 14.078 & 16.129\\
            & 15 & 13.096 & 13.436 & 15.778\\
            \hline
            TBL  & 5 & 21.93 & 22.378 & 28.33 \\
            & 10 & 21.48 & 21.972 & 44.59 \\
            & 15 & 21.776 & 27.515 & 44.072 \\
            \hline
        \end{tabular}
        \caption{Reconstruction $RRMSE$ (\%) values when repairing the test case dataset variants with 20\%, 40\% and 60\% of gaps by performing an initial reconstruction using zeros and using gappy SVD with 5, 10 and 15 retained SVD modes.}
        \label{tab:gappyzeros}
    \end{table}

When comparing the reconstruction $RRMSE$ scores from Tab. \ref{tab:rrmsegappy}, with initial reconstruction by applying linear interpolation, and Tab. \ref{tab:gappyzeros}, with initial reconstruction with zeros, for the three-dimensional cylinder at Re $= 280$, it is clear that the final reconstruction improves when linear interpolation is used performed for the initial reconstruction, since data is created based on the existing data and said data may be used during the HOSVD loop. On the other hand, for both turbulent data cases, the reconstruction $RRMSE$ scores are similar, given that, in both cases, the noise that is reconstructed via interpolation is then filtered out during the HOSVD loop. Therefore, there is no appreciable difference between using zeros or using linear interpolation to fill in the missing data. 

\subsection{Data resolution enhancement applying the SVD superresolution algorithm\label{sec:SVDsupresults}}

The SVD superresolution algorithm has been applied to the test cases described in Sec. \ref{sec:database} by enhancing the resolution of an initial dataset from half, a fourth, eighth and sixteenth of the original data resolution. The initial reconstruction is completed using linear interpolation, and for the HOSVD loop 5, 10 and 15 retained SVD modes were selected to study the reconstruction limitations when retaining a decreasing number of modes and, therefore, limited information about the data. Therefore, for each test case, a total of 12 cases are studied.

Table \ref{tab:svdsupcases} gathers the reconstruction $RRMSE$ scores for all performed cases on each test case dataset. 

\begin{table}[H]
        \centering
        \begin{tabular}{|l  l  l  l  l  l|}
        \hline
        \rowcolor{Gray}
        \hline
        \textbf{Datasets} & \textbf{modes} & \textbf{1/2} & \textbf{1/4} & \textbf{1/8} & \textbf{1/16}
        \\ \hline \hline
            Cyl3D Re280 & 5 & 11.028 & 17.894 & 28.106 & 33.926 \\
            & 10 & 8.147 & 16.48 & 28.452 & 34.722 \\
            & 15 & 5.761 & 16.031 & 28.45 & 34.821  \\
            \hline
            Cyl2D Re4000  & 5& 16.919 & 16.749 & 17.667 & 24.238 \\
            & 10 & 17.345 & 14.438 & 15.784 & 23.388 \\
            & 15 & 15.921 & 13.585 & 15.051 & 23.071 \\
            \hline
            TBL & 5 & 24.013 & 22.298 & 24.016 & 27.884  \\
            & 10 & 22.924 & 21.23 & 23.535 & 28.04  \\
            & 15 & 19.791 & 20.149 & 23.266 & 28.152  \\
            \hline
        \end{tabular}
        \caption{Reconstruction $RRMSE$ (\%) scores when enhancing the resolution of under-resolved versions of the test cases from 1/2, 1/4, 1/8 and 1/16 of the total data points when applying the SVD superresolution algorithm using 5, 10 and 15 retained SVD modes and performing an initial reconstruction using linear interpolation.}
        \label{tab:svdsupcases}
    \end{table}

\begin{figure}[H]
	\begin{center}
    \includegraphics[width=\textwidth]{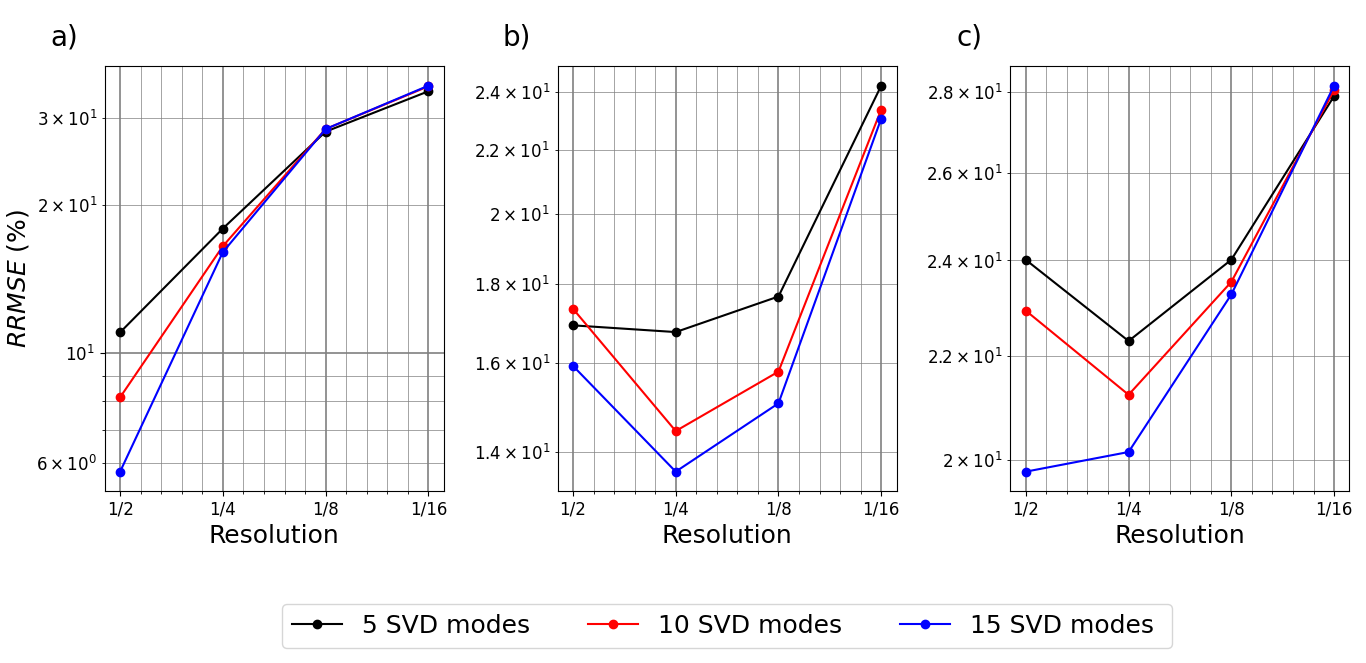}
\vskip-0.1cm
\end{center}
\vskip-0.5cm
	\caption{Reconstruction $RRMSE$ evolution for each case when enhancing the data resolution from 1/2, 1/4, 1/8 and 1/16 of the original dataset resolution to the real resolution, and 5, 10 and 15 retained singular values: a) presents the results for the three-dimensional cylinder at Re $= 280$, b) the turbulent Re $= 4000$ cylinder, and c) the turbulent boundary layer. \label{fig:rrmsesvdsup}}
\end{figure}

A clearer representation of these results can be seen in Fig. \ref{fig:rrmsesvdsup}.

For the three-dimensional cylinder at Re $= 280$ (Fig. \hyperref[fig:rrmsesvdsup]{\ref*{fig:rrmsesvdsup}(a)}), the reconstruction $RRMSE$ evolves as expected. Since the data is laminar, the $RRMSE$ decreases when the number retained of SVD modes increases, meaning more relevant information is used for reconstruction. It also shares a relationship with the resolution, since the lower the resolution, the higher the $RRMSE$, since more data needs to be generated to enhance the data resolution to the objective resolution. For the turbulent Re $= 4000$ cylinder (Fig. \hyperref[fig:rrmsesvdsup]{\ref*{fig:rrmsesvdsup}(b)}) and the turbulent boundary layer (Fig. \hyperref[fig:rrmsesvdsup]{\ref*{fig:rrmsesvdsup}(c)}), both consisting of turbulent data, they are reconstructed without the noise. Therefore, the reconstruction $RRMSE$ is higher when comparing the original data with the reconstruction. Similar to the laminar three-dimensional cylinder, the $RRMSE$ decreases when the number of retained SVD modes is increased, but the relationship between the $RRMSE$ and resolution seems to be more complex. Notice in Fig. \hyperref[fig:rrmsesvdsup]{\ref*{fig:rrmsesvdsup}(b)} that, for 10 and 15 retained SVD modes, when the initial resolution is 1/4 and 1/8, the $RRMSE$ is lower than for an initial resolution of 1/2. This occurs since the data to be generated is higher and, therefore, gives the SVD superresolution algorithm more leverage to create new data with high uncertainty (and replicate noise) which, in this case, is more similar to the original data. Also, since the number of retained modes is larger, more important information is made available during reconstruction, and some of this could be unfiltered noise. When the resolution is 1/2, the generated data is cleaner, so it does not match the noise in the original data illustrated in Fig. \ref{fig:vki4000}. 

\begin{figure}[H]
	\begin{center}
    \includegraphics[width=\textwidth]{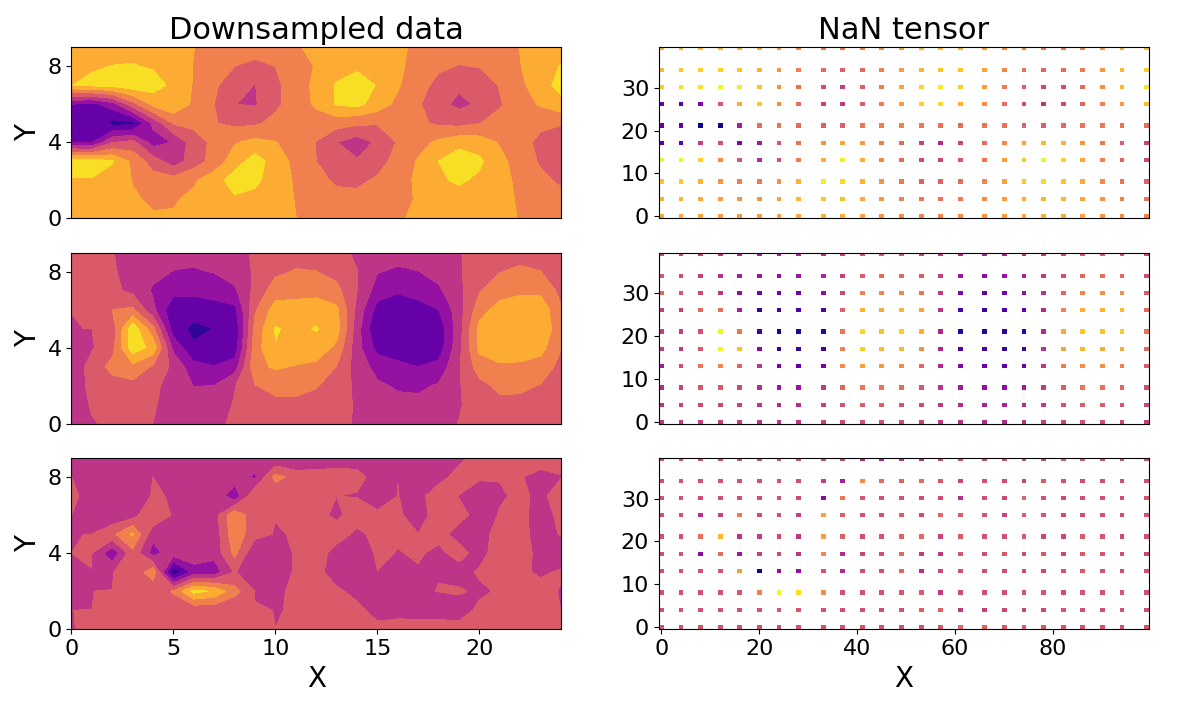}
\vskip-0.1cm
\end{center}
\vskip-0.5cm
	\caption{Data preprocessing of the three-dimensional cylinder at Re $= 280$ with 1/4 of the original resolution. From left to right and top to bottom: the under-resolved data and said data fitted on the $NaN$ tensor for the streamwise, normal and spanwise velocities. \label{fig:preprocess}}
\end{figure}

Figure \ref{fig:preprocess} shows the data preprocessing that the under-resolved dataset is submitted to (steps 1 and 2 of Sec. \ref{sec:superres}). In this case, the dataset is the three-dimensional cylinder at Re $= 280$ with a resolution of 1/4 the original data. 

After the data preprocessing stage, the initial reconstruction is performed using linear interpolation to fill in the missing data from the $NaN$ tensor.

\begin{figure}[H]
	\begin{center}
    \includegraphics[width=\textwidth]{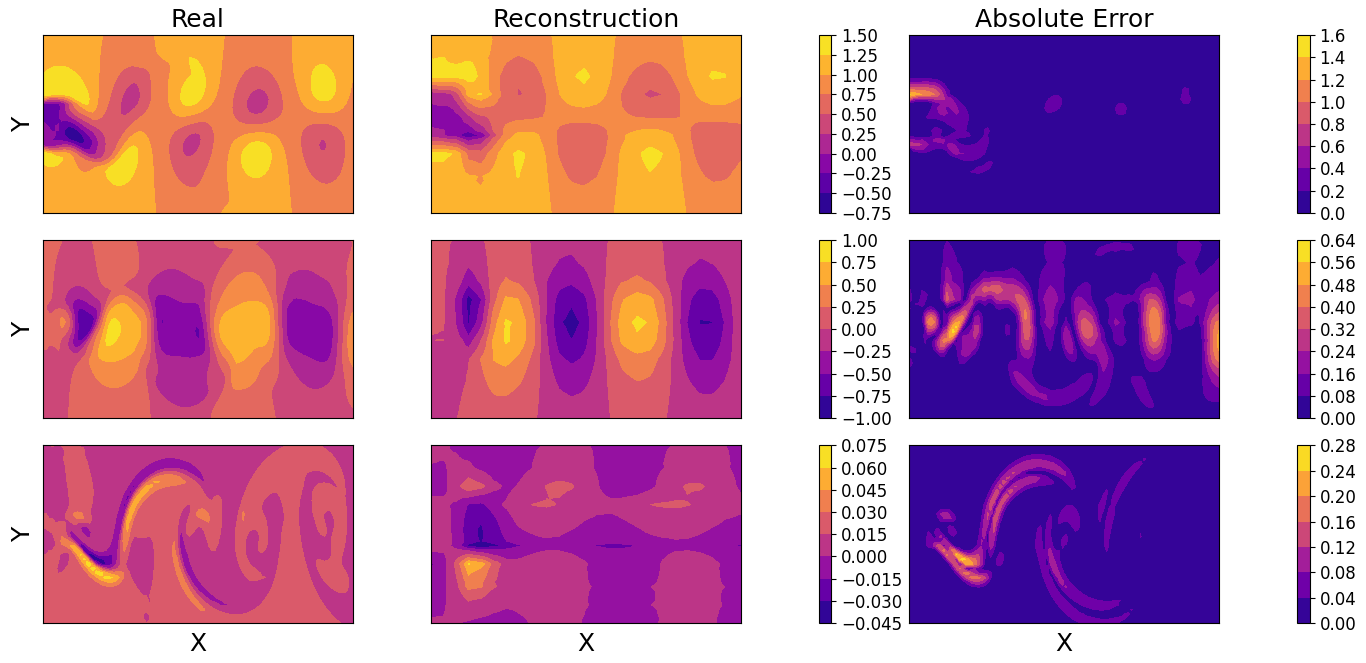}
\vskip-0.1cm
\end{center}
\vskip-0.5cm
	\caption{Resolution enhancement results for the three-dimensional cylinder at Re $= 280$ after applying the SVD superresolution algorithm to this same dataset with 1/4 the resolution of the original resolution. From left to right and top to bottom: the real data, reconstruction and absolute error for the streamwise, normal and spanwise velocities. \label{fig:svdsupperes_cyl_4_5}}
\end{figure}

The resolution enhancement results for the three-dimensional cylinder at Re $= 280$ velocity components when up-scaling the resolution from 1/4 of the real data resolution using 5 retained SVD modes can be seen in Fig. \ref{fig:svdsupperes_cyl_4_5}. The solution is presented for the highest absolute error plane and snapshot of each velocity component.

The methodology is similar to gappy SVD. The initial reconstruction is performed upon the $NaN$ tensor from Fig. \ref{fig:preprocess}, where the 1/4 of available data is evenly distributed, which is similar to applying gappy SVD to a dataset with 75\% of uniformly distributed gaps. Since, in this case, the spacing between data points is large, and not sparse like gappy SVD, this makes interpolation more difficult given the lack of neighbouring points. When a data point does not have any surrounding data and, therefore, interpolation is not possible, it is assigned the mean value of the velocity component for that snapshot.

\begin{figure}[H]
	\begin{center}
    \includegraphics[width=\textwidth]{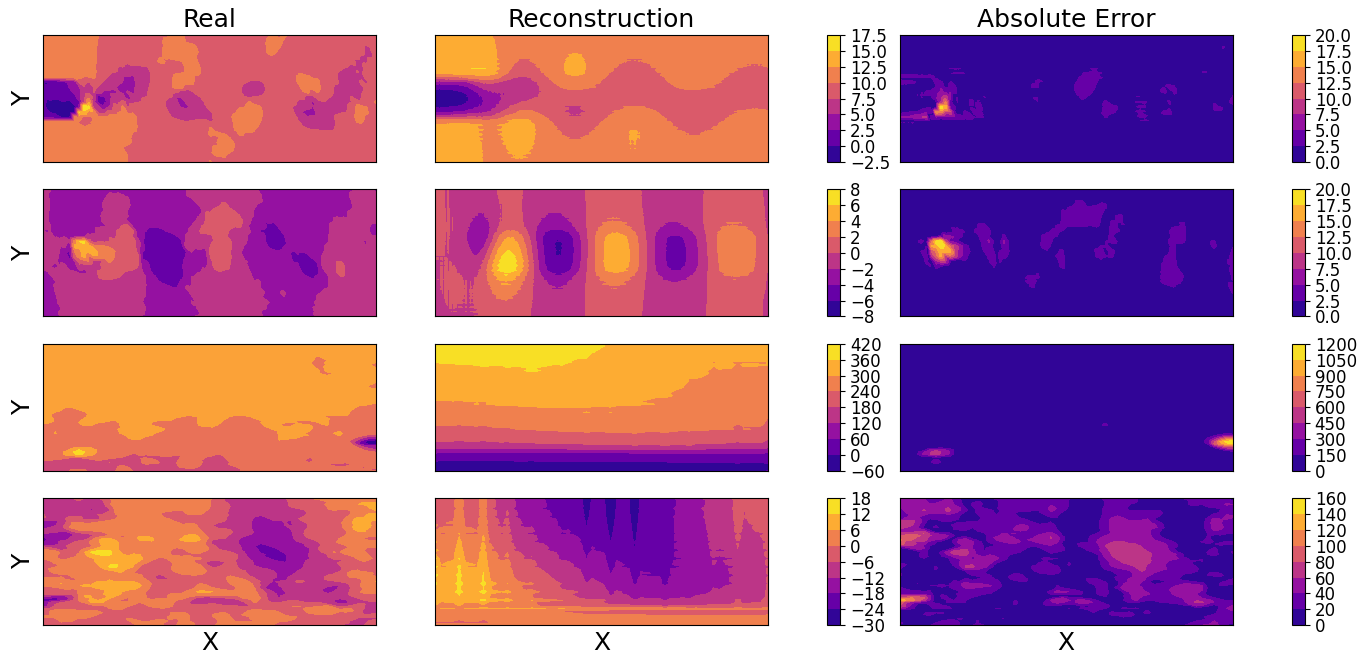}
\vskip-0.1cm
\end{center}
\vskip-0.5cm
	\caption{Resolution enhancement results after applying the SVD superresolution algorithm. From left to right and top to bottom: the real data, reconstruction and absolute error for the highest error snapshot of the streamwise and normal velocities of the turbulent Re $= 4000$ cylinder, shown in the first two rows, and the streamwise and normal velocities of the turbulent boundary layer, shown in the last two rows.  \label{fig:svd_superres_rest}}
\end{figure}

The resolution enhancement results for both the turbulent Re $= 4000$ cylinder and turbulent boundary layer using datasets 1/4 of the real resolution and retaining 5 SVD modes during reconstruction, for the highest absolute error snapshot for each velocity component, are illustrated in Fig. \ref{fig:svd_superres_rest}.

Notice that the absolute error values for each velocity component of both test cases similar or larger to the velocity field scale. These peak error values can are caused by two main reasons. First, the data enhancement in being performed using a dataset with a resolution that is 1/4 of the real data resolution. This means that, when the data is projected onto the $NaN$ tensor, the reconstruction is being performed using only 25\% of data. In addition, HOSVD smoothens the data (see the reconstruction column of Fig. \ref{fig:svd_superres_rest}), eliminating all existing noise by reconstructing the dataset using a low number of retained SVD modes. This makes the error values spike when the original data (with noise) is compared to the reconstruction (smooth), also affecting the reconstruction $RRMSE$ score.

\begin{figure}[H]
	\begin{center}
    \includegraphics[width=\textwidth]{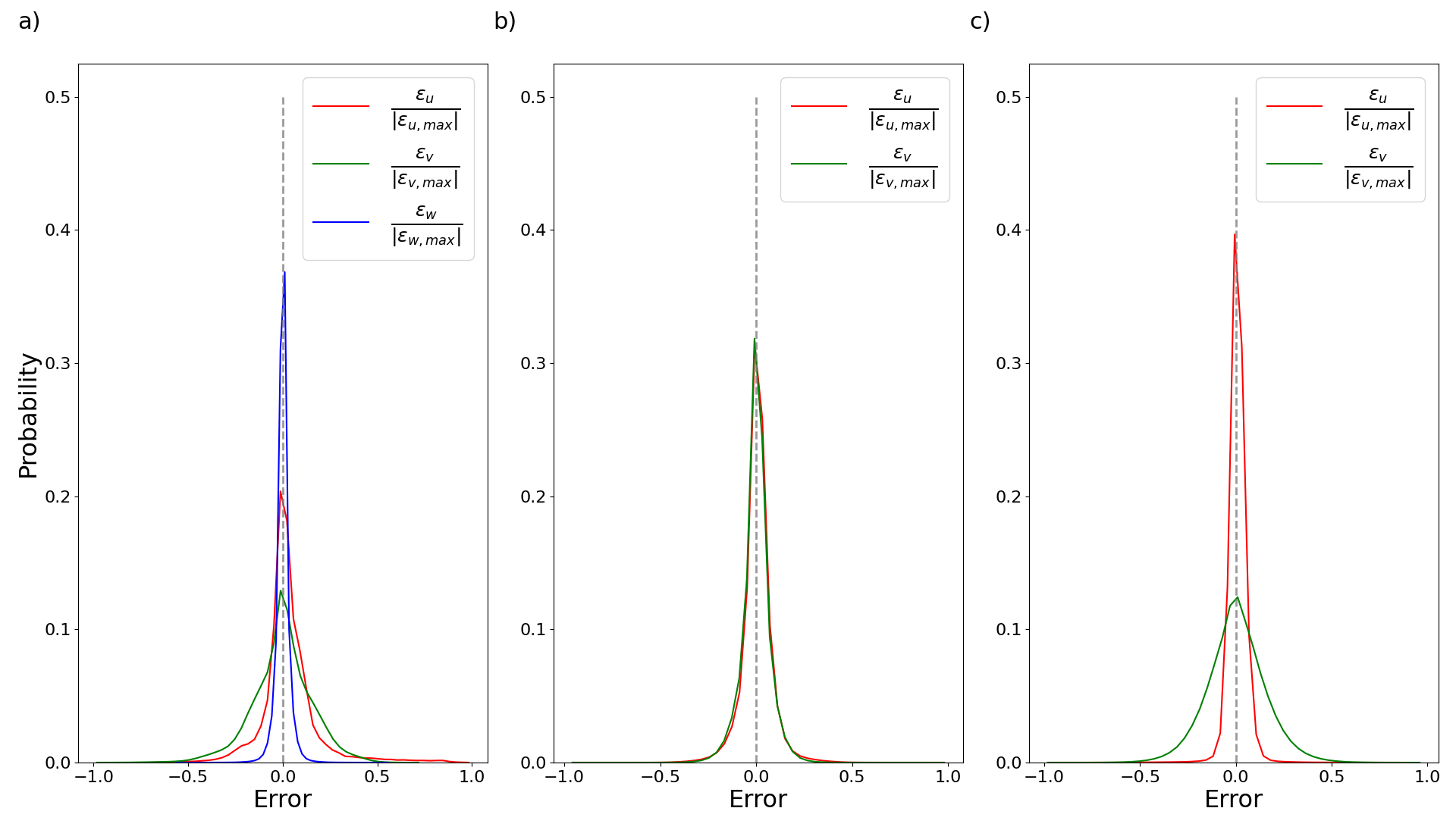}
\vskip-0.1cm
\end{center}
\vskip-0.5cm
	\caption{SVD superresolution algorithm reconstruction error probability density functions for each velocity component of: a) the three-dimensional cylinder at Re $= 280$, b) turbulent Re $= 4000$ cylinder, and c) turbulent boundary layer dataset, all for 1/4 of the real resolution. \label{fig:pdf_svd_superres}}
\end{figure}

Figure \ref{fig:pdf_svd_superres} presents the probability density functions for the reconstruction errors of all velocity components of each test case.

The reconstruction error probability density functions of all three velocity components of the three-dimensional cylinder at Re $= 280$ (Fig. \hyperref[fig:pdf_svd_superres]{\ref*{fig:pdf_svd_superres}(a)}) represent a normal distribution, with the difference between the curves laying in their width. The normal velocity reconstruction error probability distribution curve is the widest, meaning that it has the highest reconstruction error uncertainty, while the spanwise velocity probability curve is the narrowest, resembling the lowest uncertainty. The reconstruction error probability curves for the velocity components of the turbulent Re $= 4000$ cylinder (Fig. \hyperref[fig:pdf_svd_superres]{\ref*{fig:pdf_svd_superres}(b)}) and the turbulent boundary layer (Fig. \hyperref[fig:pdf_svd_superres]{\ref*{fig:pdf_svd_superres}(c)}) also represent a normal distribution. The curves of the turbulent Re $= 4000$ cylinder are close to being considered identical, while those of the turbulent boundary layer are different, since the normal velocity reconstruction error probability distribution resembles a larger uncertainty given its wider shape.

\subsection{Data resolution enhancement with deep learning superresolution \label{sec:DLsupresults}}

The results generated by the DL superresolution model, which is hybrid since it combines modal decomposition (SVD) and deep learning, are different to those presented in the previous subsection (Sec \ref{sec:SVDsupresults}), since the model adapts to the fed in data by using the hyperparameter tuner from the \textit{TensorFlow} Python library \cite{tuner}. This allows the model to use the optimal hyperparameters to train for each test case dataset. Also, in the SVD stage of this model, all modes of the under-resolved dataset are retained. It is important to note that this model is capable of forecasting high resolution data predictions based on the low resolution input data. For this reason, the reconstruction results are computed over the \textit{test data} generated from the train-test split shown in \ref{fig:ML2}, and not over the entire dataset. This has a series of effects over when computing the error. First, since the model adapts to the data, error values will tend to be in the lower range. On the other hand, since the error is being computed over a reduced amount of data, to be exact, over 20\% of the entire test case datasets, large local error values will generate a spike in the reconstruction $RRMSE$, which evaluates the global reconstruction error.

Table \ref{tab:dlrrmse} presents the reconstruction $RRMSE$ scores obtained after enhancing the resolution of the test data. 

\begin{table}[H]
        \centering
        \begin{tabular}{|l  l  l l l|}
        \hline
        \rowcolor{Gray}
        \hline
        \textbf{Datasets} & \textbf{1/2} & \textbf{1/4} & \textbf{1/8} & \textbf{1/16}
        \\ \hline \hline
            Cyl3D Re280 & 11.559 & 12.261 & 7.669 & 9.021\\
            
            Cyl2D Re4000 & 17.551  & 18.759 & 11.08 & 10.762\\
            
            TBL & 23.243 & 18.313 & 19.845 & 24.313\\
            \hline
        \end{tabular}
        \caption{Reconstruction $RRMSE$ (\%) values when enhancing the dataset resolution from 1/2, 1/4, 1/8 and 1/16 to the original resolution using the DL superresolution model.}
        \label{tab:dlrrmse}
    \end{table}



The reconstruction $RRMSE$ scores obtained applying DL superresolution remain practically in the same range for all cases. This error consistency is due to the model adapting to the data via the use of the optimal hyperparameter tuning.

\begin{figure}[H]
	\begin{center}
    \includegraphics[width=\textwidth]{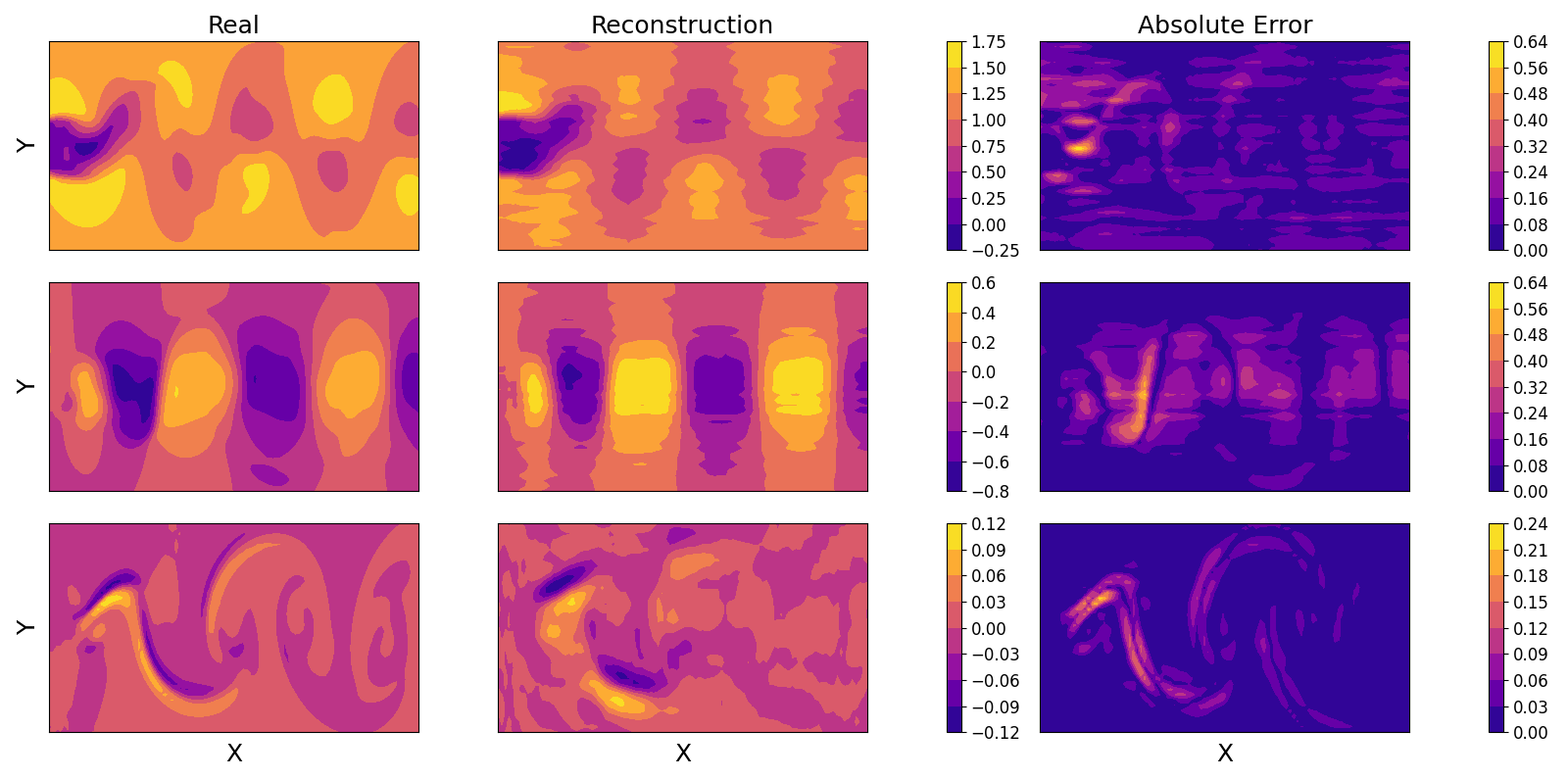}
\vskip-0.1cm
\end{center}
\vskip-0.5cm
	\caption{DL superresolution reconstruction results for the three-dimensional cylinder at Re $= 280$. From left to right and top to bottom: the real test data, the reconstruction and the absolute error of the streamwise, normal, and spanwise velocities, for the highest absolute error snapshot of each velocity component. \label{fig:dlsuperres_cyl3d}}
\end{figure}

Figure \ref{fig:dlsuperres_cyl3d} illustrates the three-dimensional cylinder at Re $= 280$ test data reconstruction results using DL superresolution. The highest absolute error snapshot is shown for each velocity component. 

The error values for all velocity component are overall low, on the scale of $10^{-1}$. Apparently, there are more data points with peak error values, which may be caused by the deep learning model or to the fact the SVD is applied in stead of HOSVD, which is more suitable for data in tensor form. The solution presented has filtered all noise, similar to the solution obtained with the SVD superresolution algorithm in Fig. \ref{fig:svdsupperes_cyl_4_5}. 

\begin{figure}[H]
	\begin{center}
    \includegraphics[width=\textwidth]{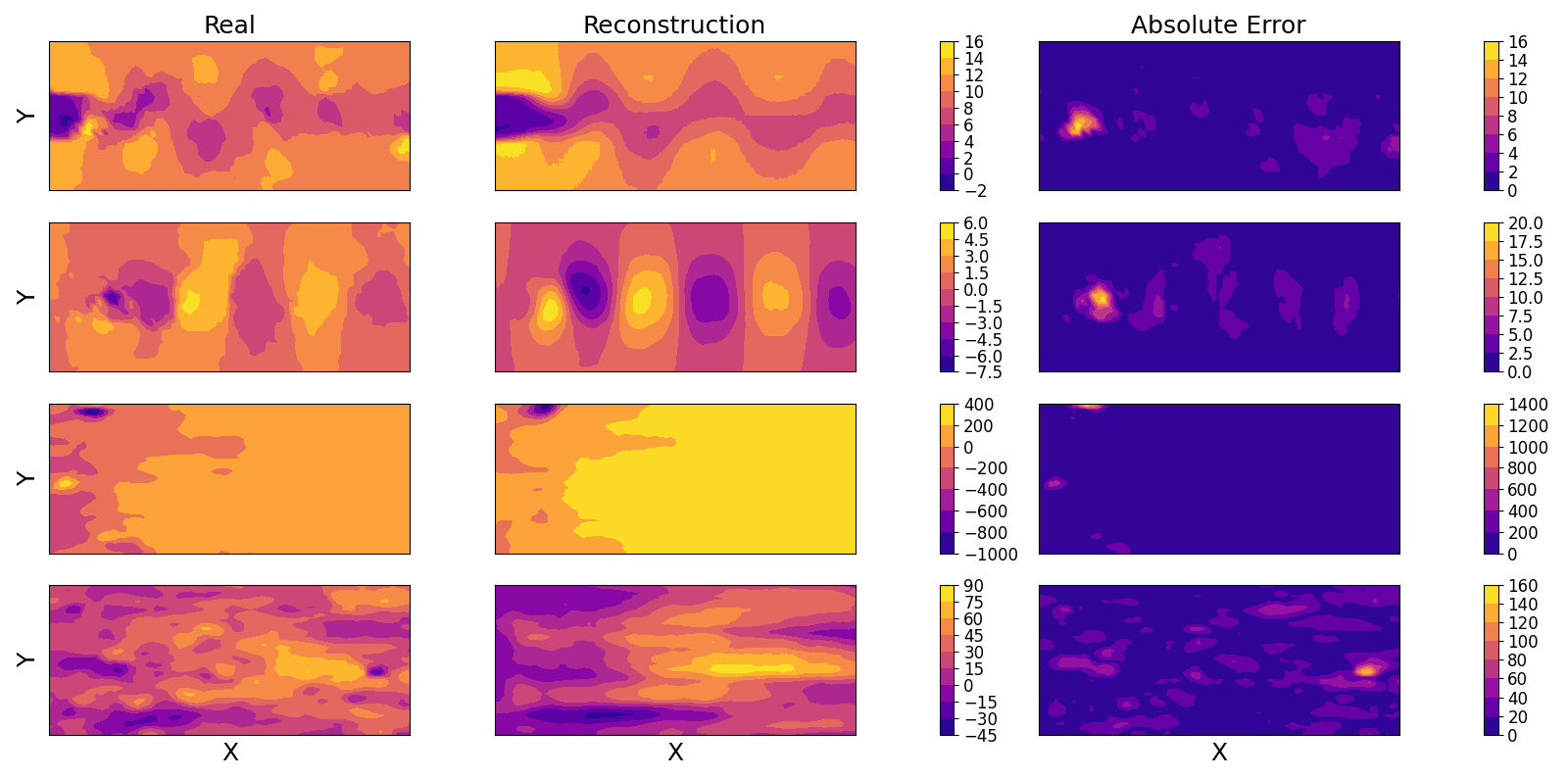}
\vskip-0.1cm
\end{center}
\vskip-0.5cm
	\caption{DL superresolution resolution enhancement results for initial resolution equal to 1/4 of the real resolution. From left to right and top to bottom: original test data, reconstruction and absolute error for the highest absolute error snapshot of the streamwise and normal velocity components of the turbulent Re $= 4000$ cylinder, and the streamwise and normal velocities of the turbulent boundary layer. \label{fig:dlsuperres_vkitbl}}
\end{figure}

The test data reconstruction results for the turbulent Re $= 4000$ cylinder and the turbulent boundary layer are shown in Fig. \ref{fig:dlsuperres_vkitbl}.

For the turbulent experimental cases, the test data reconstruction results also lack noise, such as those calculated with the SVD superresolution algorithm in Fig. \ref{fig:svd_superres_rest}, which is a consequence of the models architecture. SVD is applied on highly complex data, and the factorized matrices are passed to the deep learning model to learn the existing patterns in the data. Since some of the essential patterns are filtered out during SVD, the reconstructed test data results are smoother. Despite this, in the majority of the reconstruction absolute error snapshots, the error values are mostly 0.

\begin{figure}[H]
	\begin{center}
    \includegraphics[width=\textwidth]{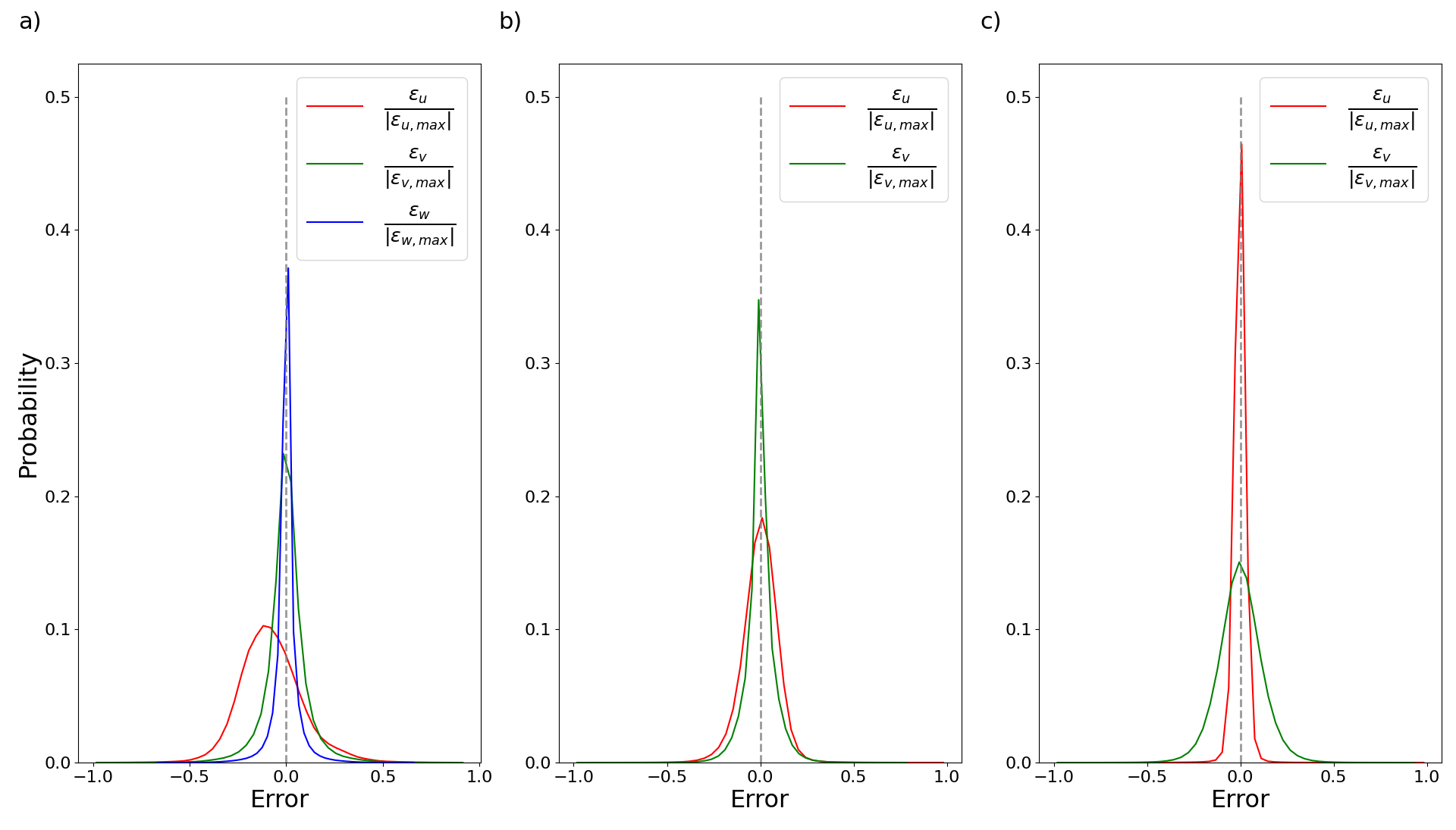}
\vskip-0.1cm
\end{center}
\vskip-0.5cm
	\caption{DL superresolution reconstruction error probability density functions for each velocity component of: a) the three-dimensional cylinder at Re $= 280$, b) turbulent Re $= 4000$ cylinder, and c) turbulent boundary layer test dataset, for the data resolution enhancement solution using an initial resolution 1/4 of the original resolution for each test dataset. \label{fig:pdfdlsuper}}
\end{figure}

The probability density functions for the velocity component reconstruction error values of the test data reconstruction of all test cases are presented in Fig. \ref{fig:pdfdlsuper}.

For the three-dimensional cylinder at Re $= 280$ (Fig. \hyperref[fig:pdfdlsuper]{\ref*{fig:pdfdlsuper}(a)}), the streamwise velocity reconstruction error follows a right skewed distribution and has a high level of uncertainty, while the normal and spanwise velocity reconstruction errors follow a normal distribution. In the case of the turbulent Re $= 4000$ cylinder (Fig. \hyperref[fig:pdfdlsuper]{\ref*{fig:pdfdlsuper}(b)}), both components follow a normal distribution, with the streamwise velocity reconstruction error having a larger level of uncertainty than the normal velocity error. For the turbulent boundary layer (Fig. \hyperref[fig:pdfdlsuper]{\ref*{fig:pdfdlsuper}(c)}), the component reconstruction errors also follow a normal distribution, with the normal velocity reconstruction error having a larger uncertainty that the streamwise velocity error. It is important to note that the data that is being analyzed is the velocity reconstruction error for the test data of each dataset, which consists of only 20\% of the total data. This means that there are less error values being calculated, so each different error value has a larger weight on the representation of the probability distribution.

\section{Conclusions\label{sec:conclusions}}

The data-driven reconstruction methods presented in this paper have proven to be effective tools to reconstruct datasets, whether the data is two- or three-dimensional, laminar or turbulent, numerical or experimental, etc. Reconstruction can be done by completing missing or corrupt data, which could be measured from a large field of sensors, or enhance the resolution of a dataset to generate a large amount of data from a reduce number of data points, which is generally used on data that has been collected from a small number of sensors. The presented methods, which are purely based on modal decomposition such as gappy SVD and the SVD superresolution algorithm, or hybrid ROMs with DL superresolution, which combines modal decomposition and deep learning, allow the reconstruction of a cleaner version of the test case datasets, since SVD and HOSVD filter out any noise that may be contained in the dataset during reconstruction. Additionally, DL superresolution has demonstrated its high reconstruction capabilities under extreme conditions, this is, when enhancing a dataset from a very low resolution. This hybrid model has also proven to be well rounded, in the sense that is can predict any type of data with a low reconstruction error thanks to the possibility of implementing hyperparameter tuning to adapt the models training parameters to the input data. This way, the model trains with the optimal hyperparameters for each case. The high reconstruction capabilities of these methods have been demonstrated in three ways. First, by presenting the reconstruction $RRMSE$ score, which provides a global estimation of how close the reconstructed data is to the real data. In addition, the absolute error contour map gives a visual understanding of how well each reconstruction algorithm performs and where the peak error values are concentrated. Finally, uncertainty quantification is used to comprehend how the error data is distributed with the range of error values and the probability of each one of these. These three error computation techniques combined give a complete and detailed understanding of how well the methods presented in this paper perform.  

\section[]{Acknowledgements}
The authors would like to express their gratitude to the research group ModelFLOWs for their valuable discussions, assistance in generating new databases, and for their support in testing some of the developed tools. The authors acknowledge the grant PID2020-114173RB-I00 funded by MCIN/AEI/ 10.13039/501100011033 and the support of Comunidad de Madrid through the call Research Grants for Young Investigators from Universidad Politécnica de Madrid. S.L.C. acknowledges the support provided by Grant No. TED2021-129774B-C21 and by Grant No. PLEC2022-009235, funded by MCIN/AEI/
10.13039/501100011033 and by the European Union
“NextGenerationEU”/ PRTR.


\bibliographystyle{elsarticle-num-names} 
\bibliography{jfmBib}





\end{document}